\documentclass[aps,a4paper,twocolumn,prl]{revtex4}
\pdfoutput=1

\usepackage{amsmath}
\usepackage{amsfonts}
\usepackage{amssymb}
\usepackage[cal=cm]{mathalfa}
\usepackage{graphicx}
\usepackage{braket}
\usepackage{sidecap}
\usepackage{color}
\usepackage{bbm}
\usepackage{nicefrac}
\usepackage{float}
\usepackage[dvipsnames]{xcolor}
\usepackage[utf8]{inputenc}
\usepackage[normalem]{ulem}
\usepackage{soul}
\usepackage{nicefrac}
\usepackage{sidecap}
\usepackage[citebordercolor=blue]{hyperref}
\usepackage{textcomp}
\usepackage{braket}
\usepackage{bbm}
\usepackage{dsfont}
\usepackage{soul,xcolor}
\setstcolor{red}

\newcommand{\stkout}[1]{\ifmmode\text{\sout{\ensuremath{#1}}}\else\sout{#1}\fi}

\definecolor{brass}{rgb}{0.71, 0.65, 0.26}
\definecolor{cgreen}{rgb}{0.0, 0.42, 0.24}
\definecolor{cadetblue}{rgb}{0.37, 0.62, 0.63}

\DeclareMathOperator{\Tr}{Tr}

\begin{document}

\title{\textcolor{black}{Experimental Test of Entropic Noise-Disturbance Uncertainty Relations} \\for Three-Outcome Qubit Measurements}

\author{\textcolor{black}{Stephan Sponar}$^{1}$}
\email{stephan.sponar@tuwien.ac.at}
\author{\textcolor{black}{Armin Danner}$^{1}$}
\author{Vito Pecile$^{1}$}
\author{Nico Einsidler$^{1}$}
\author{\textcolor{black}{B\"ulent Demirel}$^2$}
\author{\textcolor{black}{Yuji Hasegawa}$^{1,3}$}
\email{yuji.hasegawa@tuwien.ac.at}
\affiliation{%
$^1$Atominstitut, TU Wien, Stadionallee 2, 1020 Vienna, Austria \\
$^2$Institute for Functional Matter and Quantum Technologies, University of Stuttgart, 70569 Stuttgart, Germany\\
$^3$Department of Applied Physics, Hokkaido University, Kita-ku, Sapporo 060-8628, Japan}

\date{\today}

\begin{abstract}
%Information-theoretic uncertainty relations, i.e., uncertainty relations formulated in terms of entropies, characterize the well known noise-disturbance tradeoff, thereby reflecting to which extend two non-commuting observables can be measured successively. 
\textcolor{black}{Information-theoretic uncertainty relations formulate the joint immeasurability of two non-commuting observables in terms of information entropies. }\textcolor{black}{The trade-off of the accuracy in the outcome of two successive measurements \textcolor{black}{manifests} in 
entropic noise-disturbance uncertainty relations.} Recent theoretical analysis predicts that projective measurements are not optimal, with respect to the noise-disturbance trade-offs. \textcolor{black}{Therefore the results in our previous letter [PRL 115, 030401 (2015)]} are outperformed by general quantum measurements. Here, we experimentally test a tight information-theoretic measurement uncertainty relation for \textcolor{black}{three-outcome positive-operator valued measures (POVM)}, using neutron spin-$\nicefrac{1}{2}$ qubits. The obtained results violate the lower bound for projective measurements as theoretically predicted.  
\end{abstract}

\maketitle

\emph{Introduction.---}%
According to the rules of quantum mechanics any single observable \textcolor{black}{or even a set} of compatible observables can be measured with arbitrary accuracy. However, \textcolor{black}{classically} \textcolor{black}{unanticipated consequences appear} when measuring non-commuting observables jointly\textcolor{black}{, either simultaneously or successively}. Heisenberg's seminal paper from 1927 \cite{Heisenberg27} predicts a lower bound on the uncertainty of a joint measurement of incompatible observables. On the other hand it also sets an upper bound on the accuracy with which the values of non-commuting observables can be simultaneously prepared. While in the past these two statements have often been mixed, they are now clearly distinguished as \emph{measurement} uncertainty and  \emph{preparation} uncertainty relations, \textcolor{black}{respectively}. 

While Heisenberg's paper only presented his idea heuristically, the first rigorously-proven  uncertainty relation for position $Q$ and momentum $P$ was provided by Kennard \cite{Kennard27} \textcolor{black}{as $\Delta(Q)\Delta(P)\geq\frac{\hbar}{2}$, in terms of \emph{standard deviations} defined} as $\Delta(A)^2=\langle\psi\vert A^2\vert\psi\rangle-\langle\psi\vert A\vert\psi\rangle^2$. In 1929, Robertson \cite{Robertson29} extended Kennard's relation to arbitrary pairs of observables $ A$ and $ B$  as
\begin{equation} \label{eq:Robertson}
	\Delta( A)\Delta( B)\geq\frac{1}{2}\vert\langle\psi\vert[ A, B]\vert\psi\rangle\vert,
\end{equation}
with the commutator $[ A, B]= A B- B A$. 

\textcolor{black} {It is widely accepted \cite{Deutsch83} (but nevertheless under discussion \cite{Ozawa03,Ozawa2019})} that the uncertainty relation as formulated by Robertson in terms of standard deviations $\Delta(A,\vert\psi\rangle)\Delta(B,\vert\psi\rangle)\geq\frac{1}{2}\vert\langle\psi\vert[A,B]\vert\psi\rangle\vert$ lacks an irreducible or state-independent lower bound, meaning it can become zero for non-commuting observables. Furthermore, the standard deviation is not an optimal measure for all states. Consequently, Deutsch began to seek a theorem of linear algebra in the form  $\mathcal U(A, B,\psi)\ge\mathcal B( A, B)$ and suggested to use (Shannon) \emph{entropy} as an appropriate measure. Note that Heisenberg's (and Kennard's) inequality $\Delta Q\Delta P\ge\frac{\hbar}{2}$ has that form, but its generalization Eq.(\ref{eq:Robertson}) 
%does not.

%An algebraic structure in the form $\mathcal U(A, B,\psi)\ge\mathcal B( A, B)$ is desirable for an uncertainty relation. 
%One way to avoid these problems is to use another measure of statistical dispersion.
Uncertainty relations in terms of entropy were introduced to solve both problems. The first \emph{entropic uncertainty relation} was formulated by Hirschman \cite{Hirschman57} in 1957 for the position and momentum observables, which was later improved in 1975 by Beckner \cite{Beckner75} and Bialynicki-Birula and Mycielski \cite{Birula75}. The extension to non-degenerate observables on a finite-dimensional Hilbert space was given by Deutsch in 1983 \cite{Deutsch83} as
\begin{equation}\label{eq:Deutsch}
H(A)+H(B)\geq -2\,{\rm{log}_2}\,{\Big(\frac{1+c}{2}\Big)},
\end{equation}
where $H$ denotes the Shannon entropy and \emph{incompatibility} $c={\rm{max}}_{i,j}\,\vert\langle a_i\vert b_j\rangle\vert$ is the maximum overlap between the eigenvectors $\vert a_i\rangle$ and $\vert b_j\rangle$ of observables $A$ and $B$, respectively. This relation was later improved by Maassen and Uffink \cite{Maassen88} yielding the well-known entropic uncertainty relation 
\begin{equation} \label{eq:Deutsch83}
H( A)+H( B)\geq -2\,{\rm log}_2\,c.
\end{equation}
% 
%where $H$ denotes the Shannon entropy and $c$ is the maximal overlap between the eigenvectors $\ket {a_i}$ and $\ket{ b_j}$ of the observables $A$ and $B$. 
Entropic uncertainty has proven to be a useful tool in entanglement witnessing~\cite{Berta2010}, complementarity~\cite{Coles2014} and in quantum information theory~\cite{NielsenChuang}. Initially, procedures to quantify error and disturbance are based on distance measures between target observables and measurements~\cite{Ozawa03,Ozawa04} or the associated probability distributions~\cite{Werner04}. More recently, interest has risen in information-theoretic measures, introduced first by Buscemi \emph{et al}.~\cite{Buscemi14}, but also in several subsequent alternative approaches~\cite{Coles15,Baek16,Schwonnek16,Barchielli18}.

\begin{figure*}[!t]
	\includegraphics[width=0.98\textwidth]{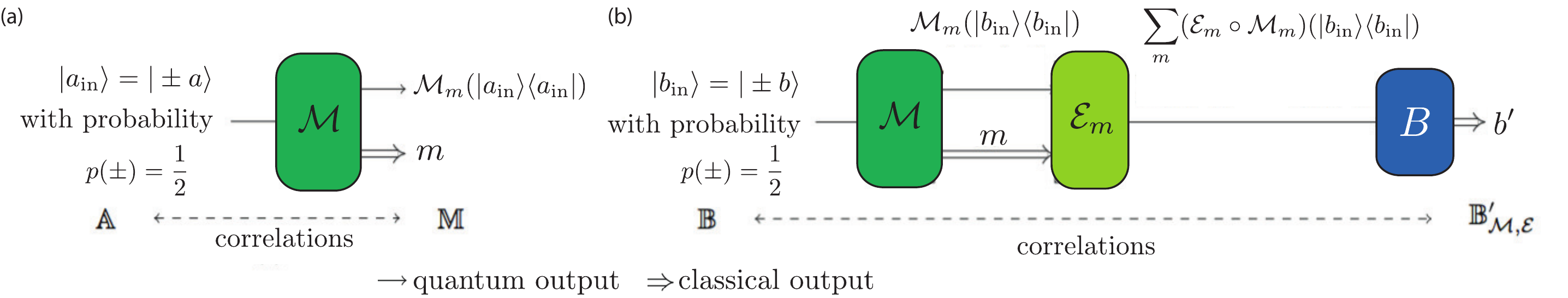}
	\caption{Schematics of the scenarios used in the information-theoretic definitions of (a) noise, $N(\mathcal M,A)$, and (b) disturbance, $D_\mathcal E(\mathcal M,B)$, \textcolor{black} {for two-level systems}. The eigenstates $\vert a_{\rm{in}}\rangle$ of $A$ (or $\vert b_{\rm{in}}\rangle$ of $B$ for disturbance) are prepared with equal probability \textcolor{black}{$p(\pm)=1/2$}, before being measured by $\mathcal M$, producing outcome $m$ and transforming the state according to $\mathcal M_m$. (b) \textcolor{black}{For the disturbance, the input states are $\ket{\pm b}$, again with probability $p(\pm)=1/2$. The result of the first measurement is classically communicated to a device applying a correction transformation $\mathcal E_m$ on the post-measurement state. The disturbance is obtained upon a subsequent projective measurement of $B$ yielding outcome $b'$ at the end.}}
	\label{fig:scheme}
\end{figure*}
\emph{Theory.---}%
To formally study measurement uncertainty relations one must define measures for two key properties of a measurement device\textcolor{black} {, more precisely  a quantum instrument, $\mathcal{M}$ \cite{Davies70,Ozawa84}} (which may in general implement an arbitrary quantum measurement with any number of outcomes): how accurately it measures a target observable $A$ (\emph{noise}), and how much it disturbs \textcolor{black}{subsequent measurements} (\emph{disturbance}).

While several definitions of noise have previously been studied theoretically and experimentally, we utilize the information-theoretic approach of~\cite{Buscemi14}, formulated as follows and schematically illustrated in Fig.\,\ref{fig:scheme}. Let $\{\ket{a}\}_a$ be the $d$ eigenstates of the $d$-dimensional target observable $A$ and \textcolor{black}{measurement device $\mathcal{M}$ being a collection $\{\mathcal M_m \}_m$ of completely positive (CP) trace-nonincreasing maps $\mathcal M_m$. The instrument $\mathcal M$ uniquely defines a} positive-operator valued measure (POVM), denoted as  ${M}=\{M_m\}_m$~\cite{NielsenChuang}. 
The noise is defined in the following scenario: the eigenstates of $A$ are randomly prepared with probability $p(a)=\frac{1}{d}$ before $\mathcal{M}$ is \textcolor{black}{applied}, producing an outcome $m$ with probability $p(m|a)=\Tr(M_m \ket{a}\!\!\bra{a})$.
If $\mathcal{M}$ accurately measures $A$ then the value of $m$ should allow one to infer $a$; if the measurement is noisy, $m$ yields less information about $a$.
This noise is quantified in terms of the conditional Shannon entropy: 
denoting the random variables associated with $a$ and $m$ as $\mathbb{A}$ and $\mathbb{M}$, respectively, the \emph{noise} of $\mathcal{M}$ \textcolor{black}{for a measurement of} $A$ is~\cite{Buscemi14} 
\begin{equation}
	N(\mathcal{M},A) = H(\mathbb{A}|\mathbb{M})=-\sum_{a,m} p(a,m)\log_2 p(a|m),
	\label{eq:noiseDefnOrg}
\end{equation}
where $p(a,m)=p(a)p(m|a)$ and $p(a|m)$ can be calculated from Bayes' theorem.

The entropic disturbance \textcolor{black}{$D(\mathcal{M}, B)$ of the apparatus $\mathcal M$ on the measurement of $B$ }is defined with respect to an analogous procedure as the noise. Uniformly distributed eigenstates $\{\ket{b_i}\}$ with eigenvalues $b_i$ associated with random variable $\mathbb{B}$ are fed to the same instrument $\mathcal{M}$ from which a post-measurement state $\rho_{m} = \mathcal{M}_{m} (\ket{b_i}\!\bra{b_i})/\text{Tr}\big(\mathcal{M}_{m} (\ket{b_i}\!\bra{b_i})\big)$ emerges. In the disturbance configuration there is an additional subsequent measurement of observable $B$ with outcomes $\{b_j'\}$. Due to the disturbing nature of the measurement apparatus $\mathcal{M}$, generally, a loss of correlation occurs. A subtle, yet important addendum to the concept of disturbance are error corrections. After measurement by $\mathcal{M}_{m_j}$, the state decomposed to the eigenstates of the measurement observables can be further transformed by a quantum operation $\mathcal{E}_{m}$ dependent on the pointer value~$m$ of the apparatus.
The disturbance  \textcolor{black}{$D_\mathcal E(\mathcal{M}, B)$} is defined as the conditional entropy $\textcolor{black}{ H(\mathbb{B}|\mathbb{B}'_{\mathcal M,\mathcal E}) }$ as
\begin{equation}
D_{\mathcal{E}}(\mathcal{M}, B) :=\textcolor{black}{ H(\mathbb{B}|\mathbb{B}'_{\mathcal M,\mathcal E}) }= -\sum_{i,j}p(b_i, b'_j) \log(p(b_i| b'_j))~.
\label{eq:DistDefn}
\end{equation}

Using these notions of noise and disturbance, for arbitrary observables $A$ and $B$ in finite-dimensional Hilbert spaces, the noise-disturbance (measurement) relation
\begin{equation}
	N(\mathcal{M}, A) + D_{\mathcal{E}}(\mathcal{M}, B) \geq -\log (\max_{i,j} |\braket{a_i|b_j}|^2)
	\label{eq:NoiseDistUR}
	\end{equation}
holds \cite{Buscemi14}.
\begin{figure}[!b]
	\includegraphics[scale=0.45]{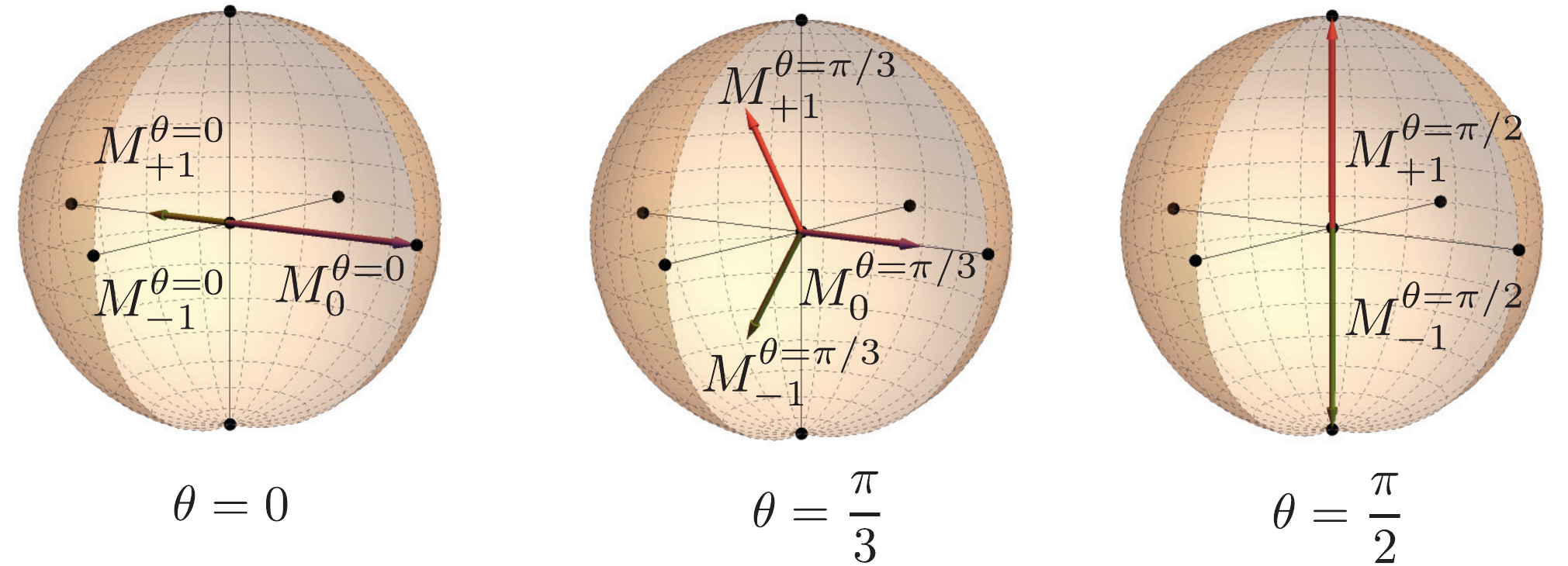}
	\caption{Bloch sphere representation of the \textcolor{black}{three-outcome} POVM $M^\theta=\{M^\theta_{-1},M^\theta_{0},M^\theta_{+1}\}$ for three selected values of the parameter $\theta$, given by $\theta=0,\pi/3,\pi/2$.
	\label{fig:POVM}}
\end{figure}
\begin{figure*}
	\includegraphics[width=0.999\textwidth]{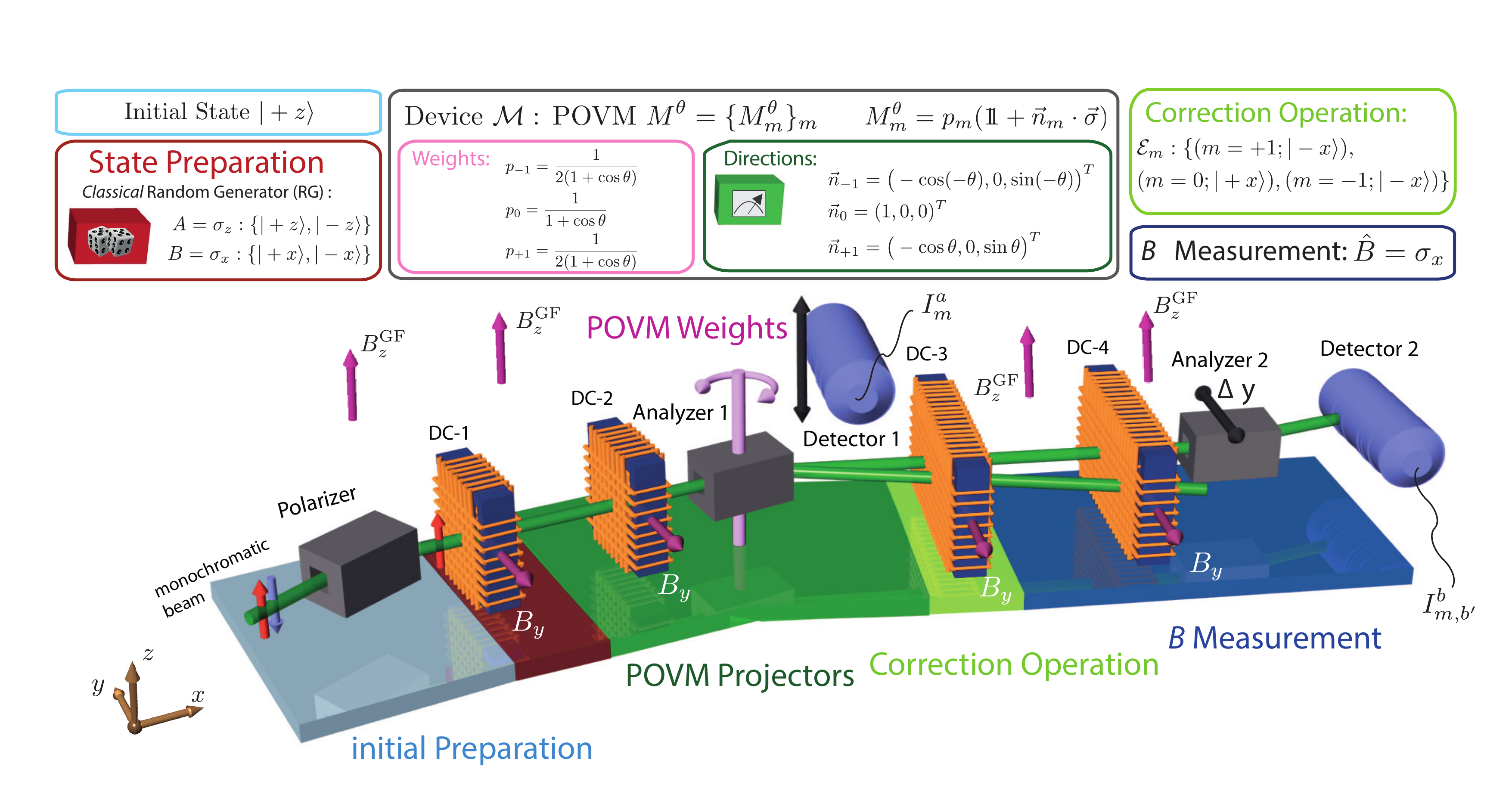}
	\caption{Schematical illustration of the neutron polarimetric setup for noise-disturbance measurement of $\mathcal M$, \textcolor{black}{representing a complete quantum instrument, consisting of \textcolor{black}{three-outcome} POVM $M^\theta$, transformation of post-measurement state (correction operation), and projective measurement $B$}. The illustration includes a \textcolor{black}{descriptive legend} of the different experimental regions. \textcolor{black}{The setup consists of three supermirror arrays (one polarizer, two analyzers), four direct current coils (DC-1,2,3,4), and two detectors.} Exploiting  Larmor  precession  of the Bloch vector around magnetic fields ($B_y,\,B_z^{\rm{GF}}$) and using supermirror arrays \textcolor{black}{to realize direction of projectors and }weights, all required spin states are prepared and the \textcolor{black}{three-outcome} POVM $M^\theta$ - as well a projective measurement $\sigma_x$ - is implemented. \textcolor{black}{Noise $N(\mathcal M,\sigma_z)$ and disturbance $D_{\mathcal{E}}(\mathcal{M}, \sigma_x)$ are evaluated from the measured intensities $I^a_m$ and $I^b_{m,b'}$, respectively}.
	\label{fig:setup}}
\end{figure*}
In \cite{Sulyok15} we experimentally tested 
\begin{equation}
	g[N(\mathcal{M}, A) ]^2+g[ D_{\mathcal{E}}(\mathcal{M}, B)]^2 \leq 1,
	\label{eq:NoiseDistUR}
	\end{equation}
where $g[x]$ is the inverse of the function $h(x)$ defined as
\begin{equation}
\textstyle h(x)=-\frac{1+x}{2}\log_2\left(\frac{1+x}{2}\right)-\frac{1-x}{2}\log_2\left(\frac{1-x}{2}\right), \,x \in [0,1].
\label{eq:BinaryEntropy}
\end{equation}
\textcolor{black}{As it turned out,} the proof given in \cite{Sulyok15} for this relation was incorrect and this relation does not hold in general, which was pointed out in \cite{Branciard16}. It should be noted that the relation does hold for projective measurements, although it can be violated by non-projective dichotomic measurements.  

The bound of Eq.(\ref{eq:NoiseDistUR}) can be violated by considering a \textcolor{black}{three-outcome} measurement $\mathcal M^\theta$ with the associated positive-operator valued measure (POVM) given by $M^\theta=\{M^\theta_{+1},M^\theta_{0},M^\theta_{-1}\}$ for $\theta\in[0,\frac{\pi}{2}]$, where 
\begin{eqnarray}\label{eq:POVM}
M^\theta_m&=&p_m(\mathds{1}+\boldsymbol{n}(\theta)_m\cdot\boldsymbol{\sigma})\quad\textrm{with \textcolor{black}{weights and directions}}\nonumber\\
p_0&=&\frac{\cos\theta}{1+\cos\theta},\quad\,\,p_{-1}=p_1=\frac{1}{2(1+\cos\theta)}\nonumber\\
\boldsymbol{n}_m&=&\big(\left(-1\right)^m\cos\left(m\,\theta\right),0,\sin\left(m\,\theta\right)\big)^T 
\end{eqnarray}
which is illustrated in Fig.\,\ref{fig:POVM} for three distinctive values of the parameter $\theta$. Note that for $\theta=\frac{\pi}{2}$ the POVM $M^\theta$ degenerates to a projective measurement in $\pm z$-direction with elements \textcolor{black}{$M^{\theta=\pi/2}_{-1}=\ket{-z}\bra{-z}$ and $M^{\theta=\pi/2}_{+1}=\ket{+z}\bra{+z}$}, \textcolor{black}{ a projective measurement of $A=\sigma_z$ resulting in zero noise}. While for $\theta=0$  the POVM element \textcolor{black}{$M^{\theta=0}_{0}$ represents the projector $\ket{+x}\bra{+x}$}, \textcolor{black}{ \emph{de facto} accounting for a projective measurement of $B=\sigma_x$, therefore zero disturbance is expected.} 

The probability of obtaining outcome $m$ when measuring a state $\rho$ is thus  \textcolor{black}{${\rm{Tr}}[M^\theta_m\rho]$. Inserting} the definition of the \textcolor{black}{three-outcome} POVM $M^\theta$ into \textcolor{black}{Eq.(\ref{eq:noiseDefnOrg}) } the noise on $A=\sigma_z$ is calculated as
\begin{equation}
N(\mathcal M^\theta,\sigma_z)=\frac{\cos\theta +h(\sin\theta)}{1+\cos\theta}.
\label{eq:Noise3}
\end{equation}
In order to determine a \textcolor{black}{lower} bound on the disturbance $D(M^\theta ,\sigma_x)$, let us consider the correction $\mathcal E^{\rm{opt}}_m$ that maps \textcolor{black}{$\boldsymbol {n}_{\pm1}$ }onto the negative $x$-axis and $\boldsymbol {n}_{0}$ onto the positive $x$-axis, respectively. Using Eq.(\ref{eq:DistDefn}) one can then calculate the joint distribution $p(b' ,b)$ and thus the upper bound on the minimum disturbance for $B=\sigma_x$ as

\begin{equation}
D_\mathcal{E}(\mathcal M^\theta,\sigma_x)=\frac{h(\cos\theta)}{1+\cos\theta}.
\label{eq:Dist3}
\end{equation}
This noise-disturbance pair from Eqs.(\ref{eq:Noise3}) and (\ref{eq:Dist3}) violates Eq.(\ref{eq:NoiseDistUR}) for all $\theta\in]0,\frac{\pi}{2}[$, which is experimentally tested here. In Sec. I of the Supplemental Material \cite{SuppPOVM} details of the theoretical framework are elaborated.

\emph{Experimental Setup.---}%
The experiment was performed at the polarimeter instrument \emph{NepTUn (NEutron Polarimeter TU wieN)} \cite{Demirel20,Demirel19,Sponar17,Demirel16,Sulyok13,Erhart12}, located at the tangential beam port of the 250\,kW TRIGA Mark II  research reactor at the Atominstitut - TU Wien, in Vienna, Austria. A schematic illustration of the experimental setup is depicted in Fig.\,\ref{fig:setup}. An incoming monochromatic neutron beam with mean wavelength $\lambda\simeq 2.02\,\AA$ ($\Delta\lambda/\lambda\simeq0.02$) is polarized along the vertical  ($+z$) direction by refraction from a \textcolor{black}{swivelling} CoTi multilayer array, \textcolor{black}{henceforth} referred to as supermirror. To prevent depolarization by stray fields, a 13 Gauss guide field $B^{\rm{GF}}_z$ pointing in the positive $z$-direction, from coils in Helmholtz configuration, is applied along the entire setup (Helmholtz coils not depicted in Fig.\,\ref{fig:setup}).

\textcolor{black}{The probability of preparation of one of the two possible initial states, that is $\ket{\pm z}$  for noise and $\ket{\pm x}$ for disturbance measurement, is determined by a classical random number generator applying one out of two possible currents in the spin rotator coil DC-1. Within the coil DC-1 a local magnetic field $B_y$, pointing in positive $y$-direction, is applied. Larmor precession around the $y$-axis is induced and the strength of $B_y$ is tuned such that it causes a spin rotation by an angle of $0$ or $\pi$ for the noise and $+\frac{\pi}{2}$ or $-\frac{\pi}{2}$  for the disturbance measurement, respectively.}

For the \textcolor{black}{three-outcome} POVM $M^\theta$ another spin rotator coil (DC-2) and the second supermirror (analyzer 1) are applied. As seen from the definition of the POVM  $M^\theta_m=p_m(\theta)(\mathds{1}+\boldsymbol{n}(\theta)_m\cdot\boldsymbol{\sigma})$, each POVM element consist of a \emph{measurement-direction} given by $\boldsymbol{n}_ m$ and a \emph{weighting} denoted as $p_m$, dependent on the parameter $\theta$. While the former is adjusted by an appropriate magnetic field strength $B_y$ in DC-2, the latter \textcolor{black}{is} set by the \textcolor{black}{horizontal} angle of refraction inside the supermirror. Note that the change in angle of the supermirror only effects the transmission (weighting) and does not change polarization of the neutrons, making this procedure a valid experimental realization of the POVM  $M^\theta$.

\textcolor{black}{For the noise-disturbance measurement the whole function of the quantum instrument has to be specified (not just the POVM it induces), which includes transformation of the post-measurement state. Consequently}, a correction operation $\mathcal E^{\rm{opt}}_m$ is applied, in order to \textcolor{black}{minimize} the disturbance $D_{\mathcal{E}}(\mathcal{M}^\theta, B)$. In our experiment $\mathcal E^{\rm{opt}}_m$ maps  \textcolor{black}{$\boldsymbol {n}_{\pm1}$} onto the negative $x$-axis and $\boldsymbol {n}_{0}$ onto the positive $x$-axis, which is achieved by Larmor precession with DC-3.

Finally, DC-4 and the third supermirror (second analyzer) perform the $B$ measurement, \textcolor{black}{which} is a simple projective measurement, where the observable is given by $B=\sigma_x$. At the end of the beam line a boron trifluoride \textcolor{black}{counting tube} (\textcolor{black}{detector 2 in Fig.\,\ref{fig:setup}}) registers all incoming neutrons. The two successively performed measurements of $M^\theta$ and $B$ result in six output intensities $I^b_{m,b'}$ for $B=\sigma_x$ (\emph{disturbance-measurement}), for each setting of $\theta$ (see \textcolor{black}{Sec.\,II} of the Supplemental Material \cite{SuppPOVM} for details of the data evaluation). For the \emph{noise-measurement} no $B$ measurement is required, thus only three output intensities $I^a_{m}$ (with $m=-1,0,1$) are obtained. 

%For the disturbance measurement $D_\mathcal E(\mathcal M^\theta,B)$ the \textcolor{red}{three-outcome} POVM measurement is followed by a subsequent projective measurement of an observable $B=\sigma_x$. In addition, a correction operation $\mathcal E^{\rm{opt}}_m$ in between the two measurements maps $\boldsymbol {n}_{-1}$ and $\boldsymbol {n}_{1}$ onto the negative $x$-axis and $\boldsymbol {n}_{0}$ onto the positive $x$-axis, respectively.

\emph{\textcolor{black}{Data treatment}.---}%
Uniformly distributed eigenstates of the observable $A=\sigma_z$, denoted as $\{\ket{a_i}\}=\{ \ket{+z},\ket{-z} \}$, are sent onto the apparatus $\mathcal M^\theta$. The correlation between the eigenvalue $a_i$ corresponding to the state prepared and the outcome $m$ measured by the apparatus $\mathcal M^\theta$, \textcolor{black}{is given by the joint probability $p(a,m)$, which in turn allows us to determine the noise.}
 %is used to determine the \emph{noise}, is characterized by the joint probability distribution $p(a,m)$. 
 The \textcolor{black}{conditional} probability $p(a\vert m)$ is then obtained via  
\begin{eqnarray}
p(a\vert m)=\frac{p(a,m)}{p(m)}=\frac{p(a,m)}{\sum_a p(a,m)},
\end{eqnarray}
allowing to calculate the noise $N(\mathcal M^\theta,A)$ using Eq.(\ref{eq:noiseDefnOrg}). The noise $N(\mathcal{M^\theta}, A)$ of the \textcolor{black}{three-outcome} POVM $M^\theta$ is determined applying the \textcolor{black}{\emph{reduced} setup; here} \textcolor{black}{an additional counting tube} (\textcolor{black}{detector 1 in Fig.\,\ref{fig:setup}) is inserted by} directly mounting it onto the exit window of the \textcolor{black}{first analyzer (second supermirror). This is done} to maintain optimal positioning, \textcolor{black}{relative to the beam,} when the supermirror, is rotated to implement the POVM weights. With this configuration a maximal count rate $I_{\rm{max}}=350$ \textcolor{black}{ccounts per second} is recorded. During the measurement the POVM parameter $\theta$ is varied between $\pi/2$ and 0 in steps of $\pi/34$ (see Sec. II of the Supplemental Material \cite{SuppPOVM} for details of the noise measurement). For each value of $\theta$ \emph{three} intensities, belonging to the POVM outputs $M_0^\theta$,  $M_{+1}^\theta$ and $M_{-1}^\theta$ are recorded in a measurement time $t_{\rm{meas}}=400$ seconds. \textcolor{black}{The conditional probability $p(a\vert m)$ is obtained via  $p(a\vert m)=I^a_{m}/\sum_{a,m} I^a_{m}$, allowing to calculate the noise $N(\mathcal M^\theta,A)$ using Eq.(\ref{eq:noiseDefnOrg}). }

\begin{figure}[!t]
	\includegraphics[scale=0.3]{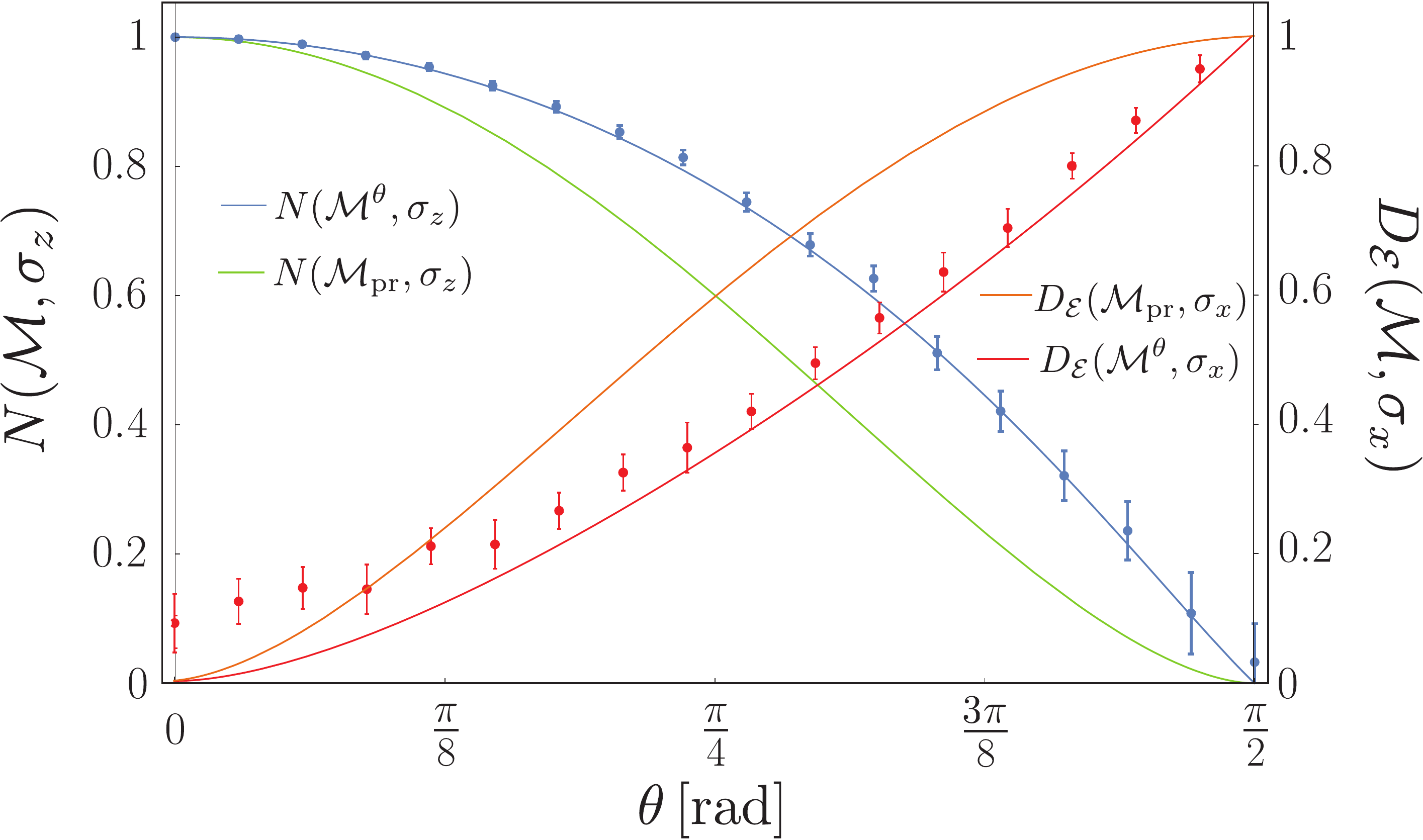}
	\caption{Plot of  noise $N(\mathcal M^\theta,\sigma_z)$ and disturbance $D_\mathcal E(\mathcal M^\theta,B)$ of the \textcolor{black}{three-outcome} POVM $M^\theta$ as a function of the POVM parameter $\theta$, together with the theoretical predictions (\textcolor{black}{blue and red} line). \textcolor{black}{For comparison, theoretical prediction of $N(\mathcal{M_{\rm{pr}}},\sigma_z)$ and $D_{\mathcal{E}}(\mathcal{M_{\rm{pr}}},\sigma_x)$ in case of projective  measurements are shown. }\textcolor{black}{Error bars correspond to plus/minus one standard deviation arising from the Poissonian statistics of the neutron count rate.}
	\label{fig:NoiseDist}}
\end{figure}

With the six conditional probabilities $p(a\vert m)$ we can calculate the noise $N(\mathcal{M^\theta},\sigma_z)$ via
\begin{eqnarray}
	N(\mathcal{M^\theta},\sigma_z)  &=& H(\mathbb{A}\vert\mathbb{M})=-\sum_{a,m}  p(a,m)\log_2 p(a|m)\nonumber\\&=& -\sum_m p(m)\sum_{a} p(a\vert m)\log_2 p(a|m),
	\label{eq:noiseDefn}
\end{eqnarray}
with $p(m)=\frac{1}{2}\textrm{Tr}[\textcolor{black}{M^\theta_m]}$. \textcolor{black}{The results of the noise measurement $N(\mathcal{M^\theta},\sigma_z)$ can be seen in Fig.\,\ref{fig:NoiseDist}. }

For the disturbance measurement $D_\mathcal E(\mathcal M^\theta,B)$ the \textcolor{black}{three-outcome} POVM measurement is followed by a subsequent projective measurement of an observable $B=\sigma_x$, \textcolor{black}{including correction operation $\mathcal E_m$ in between the two measurements.} \textcolor{black}{With this configuration a maximal count rate $I_{\rm{max}}=25$ counts per second is recorded.}
%In addition, a  maps $\boldsymbol {n}_{-1}$ and $\boldsymbol {n}_{1}$ onto the negative $x$ axis and $\boldsymbol {n}_{0}$ onto the positive $x$ axis, respectively. 
%
Uniformly distributed eigenstates of the observable $B$, denoted as $\{\ket{b_i}\}=\{\ket {+x},\ket{-x}\}$,  associated with random variable $\mathbb{B}$ are fed to the same instrument $\mathcal{M^\theta}$. \textcolor{black}{Due to the disturbing nature of the measurement apparatus $\mathcal{M^\theta}$, generally (unless $\mathcal M$ and $B$ measurement are the same), a loss of correlation occurs}. The correlation between the eigenvalue $b$ corresponding to the state prepared and the outcome $b'$
% of the second now \emph{projective} measured, 
\textcolor{black}{ corresponding to the measured eigenvalue of the projective (second) measurement yields the probabilities $p(b',b)$ which will be used to calculate the disturbance $D_{\mathcal E}(\mathcal M^\theta,B)$ via Eq.(\ref{eq:DistDefn}).}
%which will be used to define the \emph{disturbance}, is characterized by the joint probability distribution $p(b,b')$, allowing to calculate the disturbance $D_{\mathcal E}(\mathcal M,B)$ using Eq.(\ref{eq:DistDefn}).
%
%In the actual experiment, the detector was placed \emph{horizontally, transversal} to the beam. This was done in order to account for the beam displacement $\Delta y\sim 10\,$mm , caused by the second super mirror, when setting the POVM weights. 
 \textcolor{black}{In the measurement procedure of the disturbance $D_\mathcal E(\mathcal M^\theta,B)$, the POVM parameter $\theta$ is again varied between $\pi/2$ and 0 in steps of $\pi/34$}. For each value of $\theta$ now \emph{six} intensities, belonging to the $+b$ and $-b$ measurement of the POVM outputs $M_0^\theta$,  $M_{+1}^\theta$ and $M_{-1}^\theta$, are recorded in a measurement time $t_{\rm{meas}}=400$ seconds (for higher statistics also a second data set with  $t_{\rm{meas}}=800$ seconds was recorded).
%which is depicted in Fig.\,\ref{fig:NoiseDistFinal} where the disturbance $D_{\mathcal{E}}(\mathcal{M}, B)$ is plotted versus the noise $N(\mathcal M,A)$ with $A=\sigma_z$ and $B=\sigma_x$.

Finally, the disturbance $D_{\mathcal{E}}({\mathcal M^\theta},\sigma_x)$ is calculated applying the four joint probabilities $p(b,b')$, obtained by summation $p(b,b')=\sum_{m=-1}^1 p(m,b,b')$, together with the marginal probabilities $p(b')=\sum_b p(b,b')$, via the conditional entropy 
\begin{eqnarray}\label{eq:distProb}
&&D_{\mathcal{E}}(\mathcal{M^\theta}, \sigma_x) := H(\mathbb{B}|\mathbb{B}') \nonumber\\ &=&-\sum_{b,b'}p(b, b') \log_2p(b| b')=-\sum_{b,b'}p(b, b') \log_2\frac{p(b,b')}{p(b')}.\nonumber\\
\end{eqnarray}
 \textcolor{black}{The experimental results of the disturbance measurement $D_\mathcal{E}(\mathcal M^\theta,\sigma_x)$ can be seen in Fig.\,\ref{fig:NoiseDist}. The values obtained for the disturbance measurement for small values of $\theta$ are slightly higher than the theoretically predicted. This is due to the fact that for small values of $p(b,b')$ in Eq.(\ref{eq:distProb}) the disturbance  $D_\mathcal{E}(\mathcal M^\theta,\sigma_x)$ is very sensitive to the input data. Unlike in the case of the noise $N(\mathcal{M^\theta},\sigma_z)$, for the disturbance certain probabilities are predicted to be zero over the entire range of $\theta$ (see Sec.\,II.2 of the Supplemental Material \cite{SuppPOVM} for details).}

\begin{figure}[!t]
\begin{center}
	\includegraphics[scale=0.45]{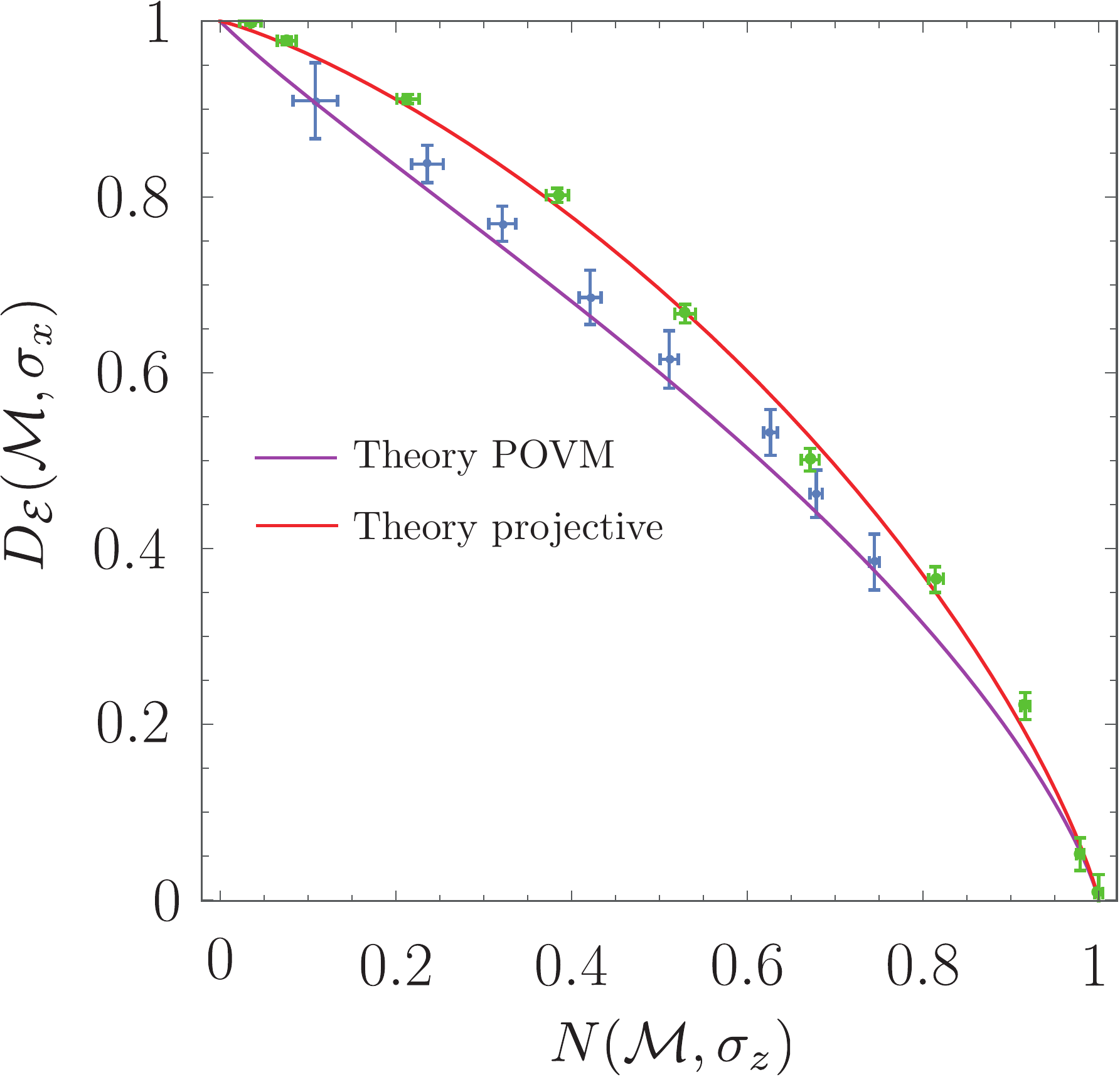}	\caption{Experimental comparison between noise-disturbance plot for \textcolor{black}{successive} projective measurements $N(\mathcal{M_{\rm{pr}}},\sigma_z)$ vs. $D_{\mathcal{E}}(\mathcal{M_{\rm{pr}}},\sigma_x)$  (green) - taken from  \cite{Sulyok15} - together with theoretical predictions in red and $N(\mathcal M^\theta,\sigma_z)$ vs. $D_\mathcal{E}(\mathcal M^\theta,\sigma_x)$ (blue)  for the \textcolor{black}{three-outcome} POVM $M^\theta$ of the measurement apparatus $\mathcal M^\theta$, \textcolor{black}{with theory} in purple. \textcolor{black}{Error bars correspond to plus/minus one standard deviation}.
	\label{fig:NoiseDistFinal}}
\end{center}
\end{figure}

\emph{\textcolor{black}{Final results}.---}%
A parametric plot of the experimental results of the noise-disturbance measurement is given in Fig.\,\ref{fig:NoiseDistFinal}, where the disturbance $D_{\mathcal{E}}({\mathcal M^\theta}, \sigma_x)$ is plotted versus the noise $N(\mathcal M^\theta,\sigma_z)$. 
Note that the final results from Fig.\,\ref{fig:NoiseDistFinal} contain disturbance measurements of $t_{\rm{meas}}=400\,$seconds, for the first 4 noise\textcolor{black}{-}disturbance pairs (\textcolor{black}{high} disturbance, \textcolor{black}{low} noise, top left), and $t_{\rm{meas}}=800\,$seconds, for the last four noise\textcolor{black}{-}disturbance pairs (\textcolor{black}{low} disturbance, \textcolor{black}{high} noise, bottom right) for better statistics.
Here, only noise\textcolor{black}{-}disturbance pairs where it is possible to \textcolor{black}{decide} whether \emph{projective} or \emph{POVM} measurements perform better (due to the size of error bars) are shown (see Sec. II of the Supplemental Material \cite{SuppPOVM} for details of the disturbance measurement).

\emph{Discussion and Outlook.---}%
In addition, Fig.\,\ref{fig:NoiseDistFinal} gives an experimental comparison with the results from the projective measurements from \cite{Sulyok15}, in terms of $N(\mathcal{M_{\rm{pr}}},\sigma_z)$ \textcolor{black}{versus} $D_{\mathcal{E}}(\mathcal{M_{\rm{pr}}},\sigma_x)$. Our experimental data clearly confirm that the \textcolor{black}{three-outcome} POVM measurement outperforms usual projective measurements, evidently reproducing the tighter bound theoretically predicted in \cite{Branciard16}.   

\textcolor{black}{At this point we want to emphasize that Fig.\,\ref{fig:NoiseDist} gives an intuitive explanation why the three-outcome POVM, defined in Eq.(\ref{eq:POVM}), outperforms projective measurements: although there is a loss comming from the noise in the POVM measurement (meaning higher noise values compared to the projective measurement), this loss is surpassed by the gain in the obtained disturbance (significantly lower disturbance values as for projective measurement). This behavior is a peculiarity of the applied three-outcome POVM. In general, increasing the number of possible outcomes has a negative (increasing) effect on the noise-disturbance bound \cite{Branciard16}. }

A next step would be investigation of two consecutive three-outcome POVM measurements. So far only the first measurement apparatus used a POVM measurement followed by a subsequent projective measurement. It is of interest to replace the projective measurement apparatus with a second  three-outcome POVM measurement and study the resulting disturbance on the second POVM measurement. 

\emph{Conclusion.---}%
We experimentally tested a tight information-theoretic measurement uncertainty relation, in terms of a proposed \textcolor{black}{three-outcome} POVM using neutron spin-$\nicefrac{1}{2}$ qubits. The obtained results of the noise-disturbance trade-off relation for three-outcome POVM outperform prior results for \emph{projective} measurements, over almost the entire measured range of the tested POVM parameter $\theta$.  

\begin{acknowledgements}\textcolor{black}{
The authors thank Alastair A. Abbott and Cyril Branciard for helpful discussions. This work was supported by the Austrian science fund (FWF) Projects No. P 30677-N36 and P 27666-N20.  }
\end{acknowledgements}

%\bibliography{/Users/stephansponar/ownCloud/General/Tex/BibTex/!myBibliography}

%%%%%%%%%%%%%%%%%%%%%% Supplementary Material  %%%%%%%%%%%%%%%%%%%%%% 

\appendix
\onecolumngrid
 
\vspace{5mm}
 
\noindent\rule{18cm}{0.4pt}
 
\section{Supplemental Material}
 \renewcommand{\theequation}{S.\,\arabic{equation}}
  \renewcommand{\thefigure}{S.\,\arabic{figure}}
 \setcounter{equation}{0}
  \setcounter{figure}{0}

%\begin{abstract}
In this supplement, we provide technical details of the data evaluation, required for determination of noise and disturbance, accompanied by the underlying theoretical framework. This complements the conceptual description given in the main text.
%\end{abstract}

%\maketitle

\section{I Theory}

The bound of Eq.(6) in the main text can be violated by \textcolor{black}{applying} a three-outcome measurement $\mathcal M^\theta$ with the associated positive-operator valued measure (POVM)  $M^\theta=\{M^\theta_{-1},M^\theta_{0},M^\theta_{1}\}$ for $\theta\in]0,\frac{\pi}{2}[$, where 
%$M^\theta_m=p_m({1\!\!1}+\boldsymbol{n}_m\cdot\boldsymbol{\sigma})=p_m\,P(\boldsymbol{n}_m)$ and $\boldsymbol{n}_m=((-1)^m\cos(m\,\theta),0,\sin(m\,\theta))^T$, with $p_0=\frac{\cos\theta}{1+\cos\theta}$ and $p_{-1}=p_1=\frac{1}{2(1+\cos\theta)}$. 
%
\begin{eqnarray}\label{eqS:POVM}
M^\theta_m&=&p_m(\mathds{1}+\boldsymbol{n}_m\cdot\boldsymbol{\sigma})\quad\textrm{with}\nonumber\\
\boldsymbol{n}_m&=&\big(\left(-1\right)^m\cos\left(m\,\theta\right),0,\sin\left(m\,\theta\right)\big)^T \quad\textrm{and}\nonumber\\
p_0&=&\frac{\cos\theta}{1+\cos\theta},\quad\,\,p_{-1}=p_1=\frac{1}{2(1+\cos\theta)}.
\end{eqnarray}
\textcolor{black}{It is worth emphasizing}  that for $\theta=\frac{\pi}{2}$ the POVM $M^\theta$ degenerates to a projective measurement in $\pm z$-direction with corresponding elements \textcolor{black}{$M^{\theta=\pi/2}_{-1}=\ket{-z}\bra{-z}$ and $M^{\theta=\pi/2}_{+1}=\ket{+z}\bra{+z}$}, while for $\theta=0$ the POVM element \textcolor{black}{$M^{\theta=0}_{0}$ represents for the projector in $+x$-direction, denoted as $\ket{+x}\bra{+x}$}. The probability of obtaining outcome $m$ when measuring a state $\rho$ is thus $\rm{Tr}[M_m\rho]$. Plugging in the three-outcome POVM $M^\theta$  from Eq.\,(\ref{eqS:POVM}) into the definition of noise
\begin{equation}
	N(\mathcal{M^\theta},A)  = H(\mathbb{A}\vert\mathbb{M})=-\sum_{a,m}  p(a,m)\log_2 p(a|m)= -\sum_m p(m)\sum_{a} p(a\vert m)\log_2 p(a|m),
	\label{eqS:noiseDefn}
\end{equation}
we calculate the noise on $A=\sigma_z$. The theoretical predictions for the conditional probabilities $p(a\vert m)$ are given by 
\begin{equation}
p(a\vert m)=\textrm{Tr}\Big[\vert a\rangle\langle a \vert \frac{M_m}{\textrm{Tr}[M_m]}  \Big]=\frac{1}{2}\big(1+ m\,a\,\sin\theta\big)\vert m\vert +(1-\vert m\vert )\cos\theta. 
\end{equation}
With the six conditional probabilities $p(a\vert m)$ we can calculate the noise $N(\mathcal{M^\theta},\sigma_z)$ via
\begin{equation}
	N(\mathcal{M^\theta},\sigma_z)  = H(\mathbb{A}\vert\mathbb{M})=-\sum_{a,m}  p(a,m)\log_2 p(a|m)= -\sum_m p(m)\sum_{a} p(a\vert m)\log_2 p(a|m),
	\label{eqS:noiseDefn}
\end{equation}
with $p(m)=\frac{1}{d}\textrm{Tr}[\textcolor{black}{M^\theta_m]}=\frac{1}{2}\textrm{Tr}[\textcolor{black}{M^\theta_m]}$\textcolor{black}{, since for qubits we have $d=2$, as}

\begin{equation}
N(\mathcal M^\theta,\sigma_z)=\frac{\cos\theta +h(\sin\theta)}{1+\cos\theta},
\label{eqS:Noise3}
\end{equation}
\textcolor{black}{with $h(x)$ defined as $h(x)=-\frac{1+x}{2}\log_2\left(\frac{1+x}{2}\right)-\frac{1-x}{2}\log_2\left(\frac{1-x}{2}\right), \,x \in [0,1]$.}

In order to determine an \textcolor{black}{lower} bound on the disturbance $D(\mathcal M^\theta ,\sigma_x)$, let us consider the correction $\mathcal E^{\rm{opt}}_m$ that maps $\boldsymbol {n}_{-1}$ and $\boldsymbol {n}_{1}$ onto the negative $x$-axis and $\boldsymbol {n}_{0}$ onto the positive $x$-axis, respectively. Using 
\begin{equation}
D_{\mathcal{E}}(\mathcal{M^\theta}, B) := H(\mathbb{B}|\mathbb{B}') = -\sum_{b,b'}p(b, b') \log(p(b| b')=-\sum_{b,b'}p(b, b') \log_2\frac{p(b,b')}{p(b')},
\label{eqS:DistDefn}
\end{equation}
 where the joint probabilities $p(b,b')$ \textcolor{black}{are} given by
\begin{eqnarray}
p(b,b')&=&\frac{1}{2}{\rm{Tr}}\Big(\sum_{m=-1}^1\mathcal E_m\big(\mathcal M_m (\vert b\rangle\langle b\vert)\big)\vert b'\rangle\langle b'\vert\Big)={\rm{Tr}}\Big(\sum_{m=-1}^1p_m\mathcal E_m\big(P(\boldsymbol n_m) (\vert b\rangle\langle b\vert)P(\boldsymbol n_m)\big) \vert b'\rangle\langle b'\vert \Big)\nonumber\\&=&\sum_{m=-1}^1 p_m\Big(\frac{1+b\,\boldsymbol e_x\cdot\boldsymbol n_m}{2}\Big)\langle b'\vert \mathcal E_m\big(P(\boldsymbol n_m)\big)\vert b'\rangle,
\end{eqnarray}
with the \emph{optimal correction} denoted as $\mathcal E^{\rm{opt}}_m\big(P(\boldsymbol n_m)\big)=\frac{{1\!\!1}+(-1)^m \boldsymbol e_x\cdot\boldsymbol\sigma}{2}$ 
\begin{eqnarray}
p(b,b')=\sum_{m=-1}^1 p_m\Bigg(\frac{1+b\,\boldsymbol e_x \cdot\boldsymbol n_m}{2}\Bigg)\Bigg(\frac{1+(-1)^m\,b'}{2}\Bigg)=\frac{1-b'+(1+b'+2bb')\cos\theta}{4(1+\cos\theta)}
\end{eqnarray}
and the marginal given by summation
\begin{eqnarray}
p(b')=\sum_b p(b,b')=\frac{1-b'+\cos\theta+b'\cos\theta}{2+2\cos\theta}.
\end{eqnarray}
\textcolor{black}{Note that our applied correction operation, that is mapping $\boldsymbol {n}_{-1}$ and $\boldsymbol {n}_{1}$ onto the negative $x$-axis and $\boldsymbol {n}_{0}$ onto the positive $x$-axis, differs from the procedure originally prosed, that leaves the state unchanged on outcome $m=0$. However, both approaches are optimal.}

Finally, the disturbance is calculated applying the four joint probabilities $p(b,b')$ from above via the conditional entropy $H(\mathbb{B}|\mathbb{B}')$, as \textcolor{black}{$D_{\mathcal{E}}(\mathcal{M}^\theta, B) := -\sum_{b,b'}p(b, b') \log_2p(b| b')=-\sum_{b,b'}p(b, b') \log_2\frac{p(b,b')}{p(b')}$.}
One can finally calculate the upper bound on the disturbance for $B=\sigma_x$ as
\begin{equation}
D_\mathcal{E}(\mathcal M^\theta,\sigma_x)=\frac{h(\cos\theta)}{1+\cos\theta}.
\label{eqS:Dist3}
\end{equation}
This noise-disturbance pair from Eqs.\,(\ref{eqS:Noise3}) and (\ref{eqS:Dist3}) violates Eq.\,(6) of the main text for all $\theta\in\,]0,\frac{\pi}{2}[\,$ which is experimentally tested here.

% A comparison between the theoretical predictions for the noise-disturbance trade-off relations for projective and the three output POVM measurement represented by $M^\theta$ is given in Fig.\,\ref{fig:ND_Theory}.

%In the case of \emph{No correction}, denoted as $\mathcal E^{\rm{No}}_m\big(P(\boldsymbol n_m)\big)=\frac{{1\!\!1}+ \boldsymbol n_m\cdot\boldsymbol\sigma}{2}$ 
%\begin{eqnarray}
%p(b,b')=\sum_{m=-1}^1 p_m\Bigg(\frac{1+b\,\boldsymbol e_x \cdot\boldsymbol n_m}{2}\Bigg)\Bigg(\frac{1+b'\,\boldsymbol e_x \cdot\boldsymbol n_m}{2}\Bigg)=\frac{1}{4}(1+b\,b'\,\cos\theta).
%\end{eqnarray}
%Since $p(b') = 1/2$ the uncorrected disturbance turns out to be just $D_{{1\!\!1}}(M^\theta,\sigma_x) = h(\cos \theta)$, which is plotted in Fig.\,\ref{fig:ND_TheoryUncorrected}.
%
%\begin{figure*}[!t]
%	\includegraphics[width=0.45\textwidth]{ImagesSupp/NDTheory}
%	\caption{Theoretical predictions of noise-disturbance plot for projective measurements $N(\mathcal{M_{\rm{pr}}},\sigma_z)$ vs $D_{\mathcal{E}}(\mathcal{M_{\rm{pr}}},\sigma_x)$ in red (taken from  \cite{Sulyok15}) and $N(M^\theta,\sigma_z)$ vs. $D_\mathcal{E}(M^\theta,\sigma_x)$ for the 3 output POVM $M^\theta$ of the measurement apparatus $\mathcal M$ in purple. }
%	\label{fig:ND_Theory}
%\end{figure*}
%

%\newpage

\section{II Data Treatment}%

\subsection{II.1 Noise Measurement}%

Uniformly distributed eigenstates of the observable $A=\sigma_z$, denoted as  $\{\ket{a_i}\}=\{ \ket{+z},\ket{-z} \}$, are sent onto the apparatus $\mathcal M^\theta$. \textcolor{black}{The correlation between the eigenvalue $a_i$ corresponding to the state prepared and the outcome $m$ measured by the apparatus $\mathcal M$ is used to determine the \emph{noise} $N(\mathcal{M^\theta}, A)$. This correlation is quantitatively characterized} by the joint probability distribution $p(a,m)$. The conditional probability $p(a\vert m)$ is then obtained via  $p(a\vert m)=\frac{p(a,m)}{p(m)}$, allowing to calculate the noise $N(\mathcal M^\theta,A)$ using Eq.\,(\ref{eqS:noiseDefn}). The noise $N(\mathcal{M^\theta}, A)$ of the \textcolor{black}{three-outcome} POVM $M^\theta$ is determined applying the \emph{reduced} setup, where 
\begin{figure*}[!b]
	\includegraphics[width=0.85\textwidth]{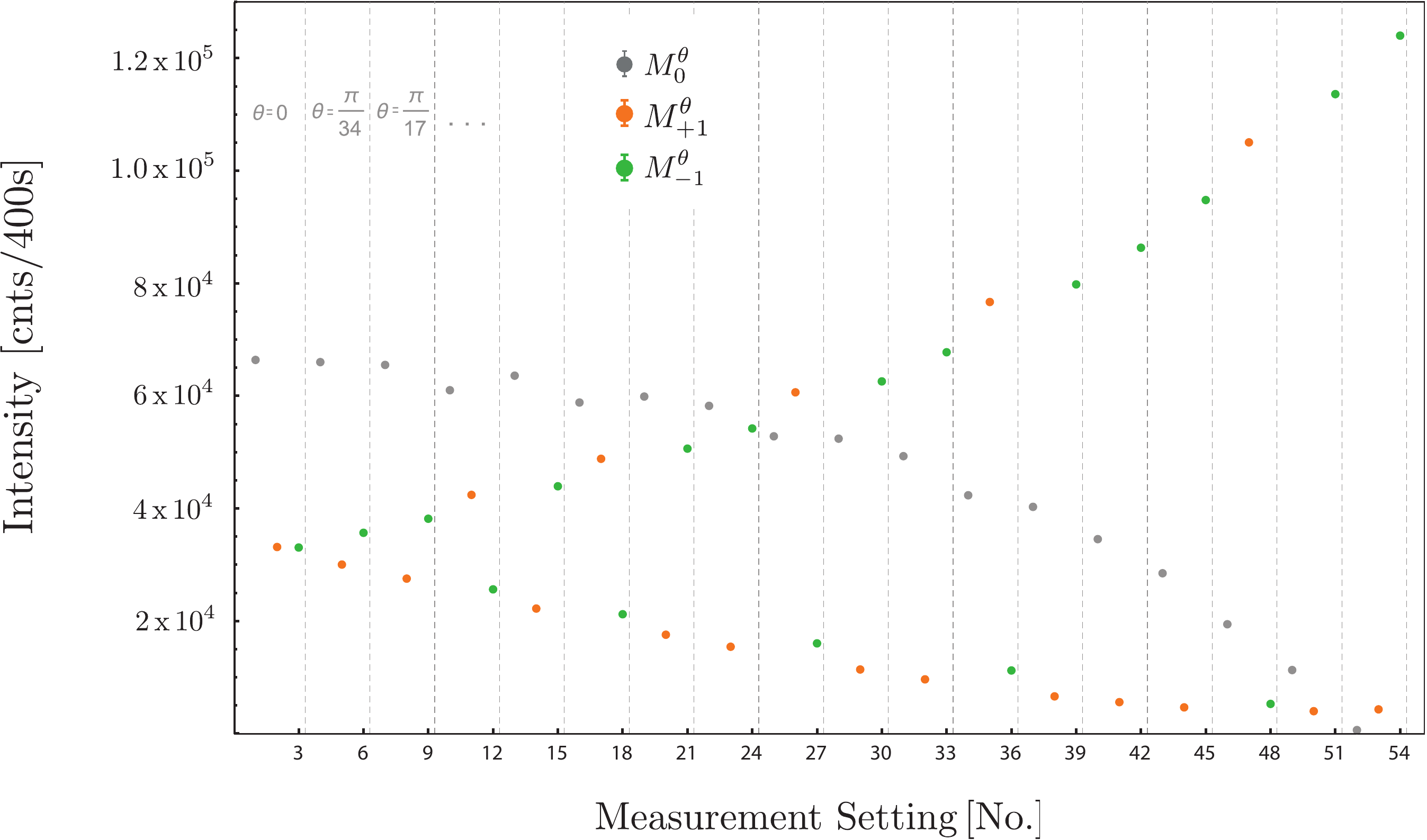}
			\caption{Raw data $I^a_m$ (with $m=-1,0,+1$) of the \emph{noise} measurement $N(M^\theta,\sigma_z)$ of the \textcolor{black}{three-outcome} POVM $M^\theta$ for a measurement time of 400 seconds. Error bars ($\pm 1$ standard deviation) are below the size of points.	
			\label{fig:DataNoise}}
\end{figure*}
the detector ($^3$He cylindric count tube, diameter \o=1 inch) \textcolor{black}{is} directly mounted onto the exit window of the second supermirror to maintain optimal positioning when the supermirror is rotated (to implement the POVM weights). With this configuration a maximal count rate $I_{\rm{max}}=350$ \textcolor{black}{counts per second} is recorded. During the measurement the POVM parameter $\theta$ is varied between $\pi/2$ and 0 in steps of $\pi/34$. 
For each value of $\theta$ \emph{three} intensities, belonging to the POVM outputs $M_0^\theta$,  $M_1^\theta$, and $M_{-1}^\theta$ (denoted as $I^a_m$ with $m=-1,0,1$) are recorded in a measurement time $t_{\rm{meas}}=400$ \textcolor{black}{seconds}, which is plotted in Fig.\,\ref{fig:DataNoise}. The particular order of the POVM elements, that is starting with $M_0^\theta$ followed by $M_{+1}^\theta$ and $M_{-1}^\theta$, has experimental reasons, namely to \textcolor{black}{reduce} the number \textcolor{black} {of} movements of the neutron optical components. %

\begin{figure*}[!t]
	\includegraphics[width=0.97\textwidth]{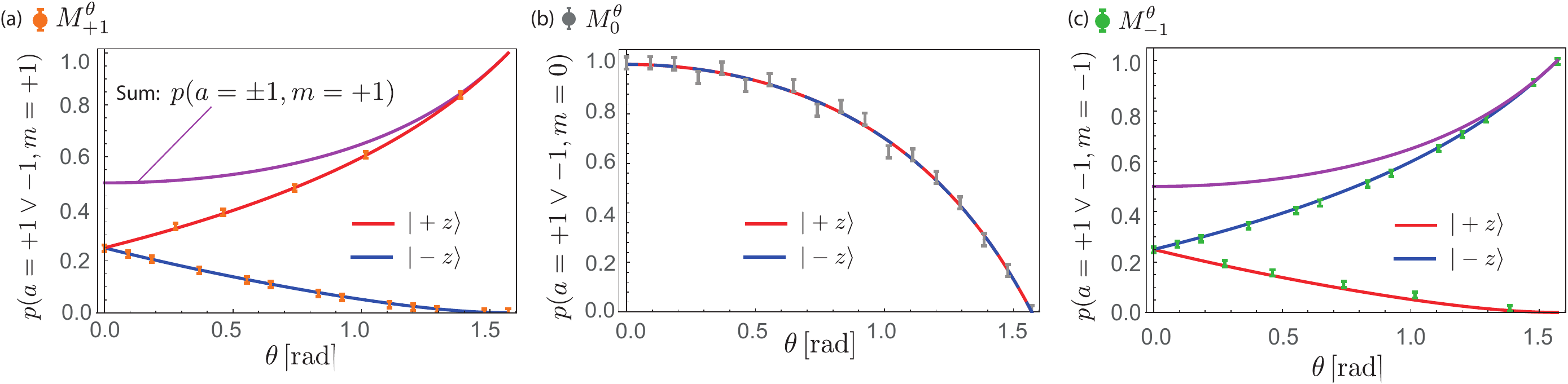}
			\caption{Normalized data of $M_{+1}^\theta, M_{0}^\theta$ and $M_{-1}^\theta$, in (a) - (c), \textcolor{black}{respectively, }with "indefinite" input state ($\vert +z\rangle$ or $\vert -z\rangle$).
			\label{fig:Norm_Pobs}}
\end{figure*}
 
For each value of $\theta$ an initial state (eigenstate of $A=\sigma_z$) is chosen by random generator. The result is \emph{blinded} during the measurement but stored in file for a later comparison with the obtained values for the noise  $N(\mathcal{M}, A)$. The following sequence was randomly generated:  
\begin{eqnarray*}
& \{\theta=0,\vert +z\rangle\},  \{\theta=\frac{\pi}{34},\vert -z\rangle\}, \{\theta=\frac{\pi}{17},\vert -z\rangle\},  \{\theta=\frac{3\pi}{34},\vert +z\rangle\},  \{\theta=\frac{2\pi}{17},\vert -z\rangle\}, \{\theta=\frac{5\pi}{34},\vert +z\rangle\}, \{\theta=\frac{3\pi}{17} ,\vert -z\rangle\} \nonumber\\
&   \{\theta=\frac{7\pi}{34},\vert -z\rangle\},    \{\theta=\frac{4\pi}{17},\vert +z\rangle\},      \{\theta=\frac{9\pi}{34},\vert -z\rangle\},     \{\theta=\frac{5\pi}{17},\vert -z\rangle\},   \{\theta=\frac{11\pi}{34},\vert +z\rangle\},  
 \{\theta=\frac{6\pi}{17},\vert -z\rangle\},   \{\theta=\frac{13\pi}{34},\vert -z\rangle\}   \nonumber\\  &   
 \{\theta=\frac{7\pi}{17},\vert -z\rangle\},  \{\theta=\frac{15\pi}{34},\vert +z\rangle\},  \{ \theta=\frac{8\pi}{17},\vert -z\rangle\},   \{\theta=\frac{\pi}{2},\vert -z\rangle\} 
  \end{eqnarray*}

\begin{figure*}[!b]
	\includegraphics[width=0.85\textwidth]{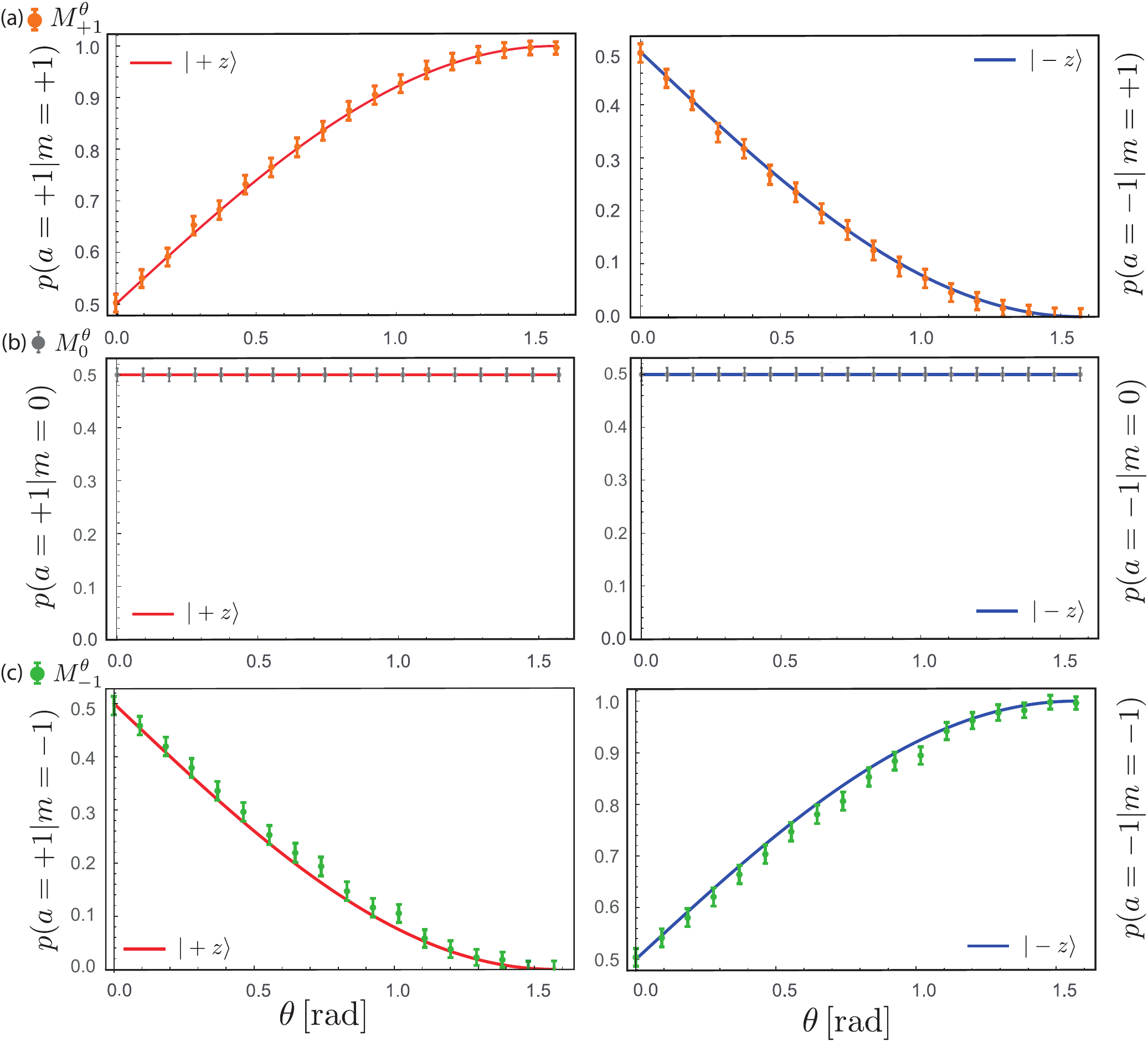}
			\caption{Conditional probability $p(a=\pm 1\vert m={-1,0,+1})$ for $\vert +z\rangle$ branch left and $\vert -z\rangle$ right.
			\label{fig:CondProbs}}
\end{figure*}

The count rates are detangled \textcolor{black}{according} to their corresponding POVM output, and data corrections are performed: First a background correction is applied, by subtraction the background counts of $I_m^{\rm{bg}}=1.37\pm 0.03\,$ \textcolor{black}{counts per second} resulting in the intensity $^{\rm{bgCorr}}I^a_m$. A second correction is performed, by taking the finite contrast for our system, measured as $C=95$\,\% into \textcolor{black}{account}. Next the count rates are normalized by the total count rate. The statistical error is given by square root of the observed count rate $\sqrt{N}$ (due to Poissonian statistics of the neutron count rates), before calculating the normalized count rate. The systematic error stems from the imperfection of the spin rotators and is estimated as $\sim 0.7$\,deg.

Normalized data of $M_{+1}^\theta, M_{0}^\theta$ and $M_{-1}^\theta$ is plotted in Fig.\,\ref{fig:Norm_Pobs}. Using the theoretical prediction of the sum of the two probabilities given by $p(a=\pm 1,m=+1)=p(a=+1,m=+1)+p(a=-1,m=+1)$, the joint probabilities $p(a=+1,m=+1)$ and p(a=-1,m=+1) can be derived for each individual value of $\theta$. 

The results for $M_{+1}^\theta$ are plotted in Fig.\,\ref{fig:Norm_Pobs} (a); apart from $\theta=0$\textcolor{black}{, where the initial states are indistinguishable,} the initial state can be inferred with a \textcolor{black}{distinctive} probability from $p(a=+1\vee a=-1,m=+1)$.  For the next output element that is  $M_0^\theta$ the situation is different. As can be seen from the normalized count rate of  $M_0^\theta$, which is plotted below in Fig.\,\ref{fig:Norm_Pobs} (b), it is impossible to infer which eigenstate of $\sigma_z$ was sent, since the theoretical \textcolor{black}{predictions} are exactly the same.  Finally, we take a look at the third output element, that is  $M_{-1}^\theta$, which is depicted in Fig.\,\ref{fig:Norm_Pobs} (c). Note that all theory curves from the output port $M_{-1}^\theta$ for input state $\vert +z\rangle$ correspond to those of  $M_{+1}^\theta$ for input state $\vert -z\rangle$. 
 Using
 \begin{equation}
p(a\vert m)=\frac{p(a,m)}{p(m)}=\frac{p(a,m)}{\sum_a p(a,m)}
\end{equation}
 the \emph{conditional probabilities} $p(a=+1\vert m=+1)$ and  $p(a=-1\vert m=+1)$ are calculated, which is depicted together with the theoretical predictions in Fig.\,\ref{fig:CondProbs} (a). The \emph{identical} data sets of $M_0^\theta$ are taken for the joint probabilities $p(a=+1,m=0)=p(a=-1,m=0)$ and for the conditional probabilities $p(a=+1\vert m=0)=p(a=-1\vert m=0)$, which is plotted in Fig.\,\ref{fig:CondProbs} (b). The conditional probabilities $p(\pm a\vert m=-1)$ are determined in analogous manner from $p(a=+1\vee-1,m=-1)$ via  $p(a=+1,m=-1)$ and $p(a=-1,m=-1)$ resulting in  $p(a=+1\vert m=-1)$ and $p(a=-1\vert m=-1)$, which is illustrated in Fig.\,\ref{fig:CondProbs} (c).

\begin{figure}[!h]
	\includegraphics[scale=0.4]{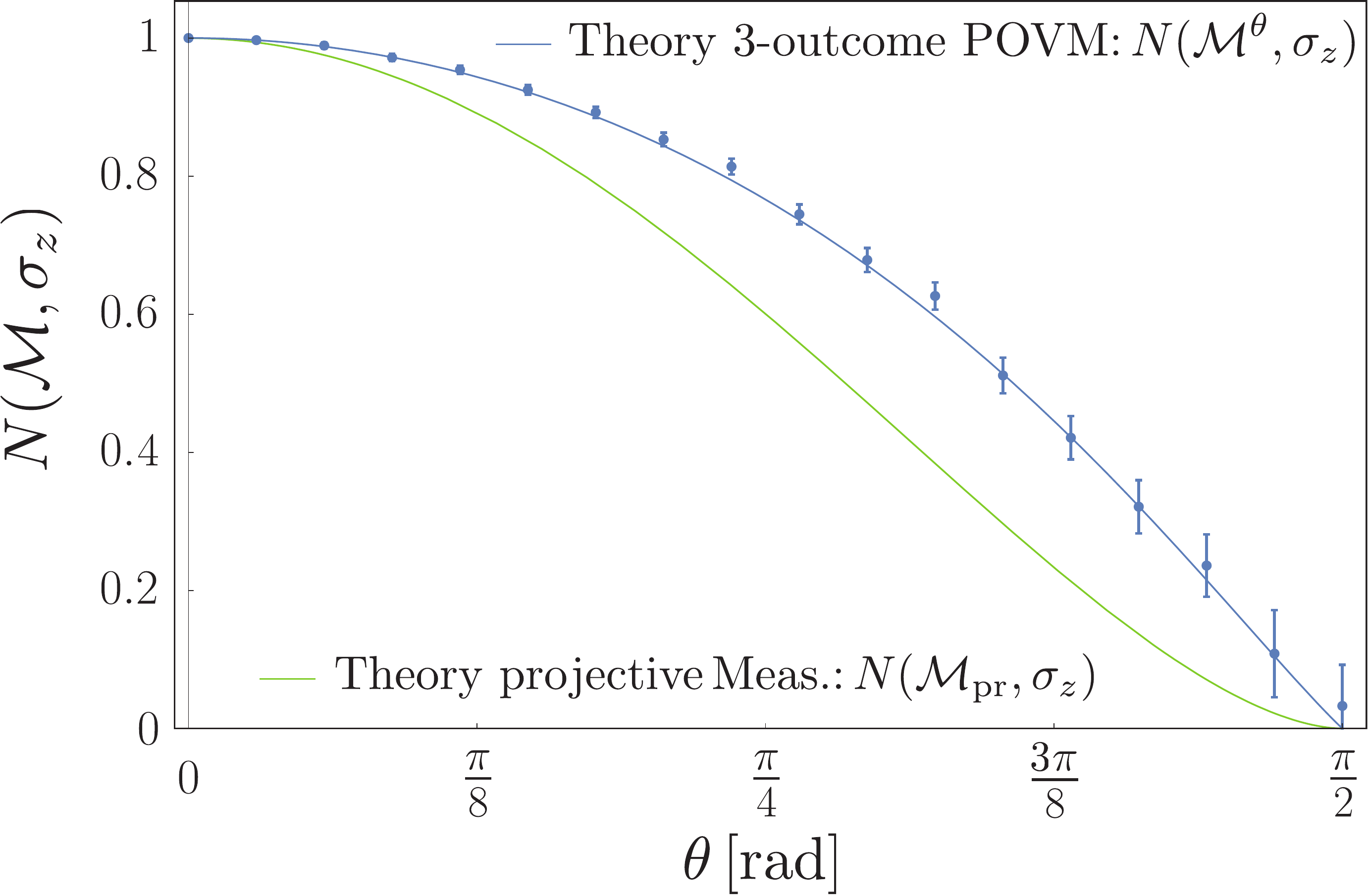}
	\caption{Plot of the noise $N(\mathcal M^\theta,\sigma_z)$ of the \textcolor{black}{three-outcome} POVM $M^\theta$ as a function of the POVM parameter $\theta$, together with the theoretical predictions for POVM and projective measurements. \textcolor{black}{Error bars correspond to plus/minus one standard deviation.}
	\label{figSupp:NoiseFinal}}
\end{figure}

 The theoretical predictions (red and blue curves in Fig.\,\ref{fig:CondProbs}) for the conditional probabilities $p(a\vert m)$ are given by 
 \begin{equation}
 p(a\vert m)=\textrm{Tr}\Big[\vert a\rangle\langle a \vert \frac{M_m}{\textrm{Tr}[M_m]}  \Big]=\frac{1}{2}\big(1+ m\,a\,\sin\theta\big)\vert m\vert +(1-\vert m\vert )\cos\theta. 
 \end{equation}
 With the six conditional probabilities $p(a\vert m)$ we can calculate the noise $N(\mathcal{M^\theta},\sigma_z)$ via
\begin{equation}
	N(\mathcal{M^\theta},\sigma_z)  = H(\mathbb{A}\vert\mathbb{M})= -\sum_m p(m)\sum_{a} p(a\vert m)\log_2 p(a|m),
	\label{eqS:noiseDefn2}
\end{equation}
with $p(m)=\frac{1}{2}\textrm{Tr}[\textcolor{black}{M^\theta_m]}$ . The final results of the noise measurement $N(\mathcal{M^\theta},\sigma_z)$, together with the theoretic prediction $N(\mathcal M^\theta,\sigma_z)=\frac{\cos\theta +h(\sin\theta)}{1+\cos\theta}$ for the three-outcome POVM measurement and $N(\mathcal M_{\rm{pr}},\sigma_z)=h(\cos\theta)$ for projective measuremnts, can be seen in Fig.\,\ref{figSupp:NoiseFinal}. 
%As seen from  Fig.\,\ref{fig:NoiseFinal}., the theoretical predictions of $N(\mathcal{M^\theta},\sigma_z)$ are reproduced evidently, indicating full experimental control of the \textcolor{red}{three-outcome} POVM over the entire range of the control parameter $\theta$

%The theoretical predictions for the noise $N(\mathcal{M^\theta},\sigma_z)$ are evidently reproduced over the entire range of the POVM parameter $\theta$. . 

%
%The theoretical predictions for the noise $N(\mathcal{M^\theta},\sigma_z)$ are evidently reproduced over the entire range of the POVM parameter $\theta$.

 %
 
%\begin{figure}[!h]
%	\includegraphics[scale=0.5]{Images/Noise_New_V4}
%	\caption{Final plot of the noise $N(M^\theta,\sigma_z)$ of the three output POVM $M^\theta$ as a function of the POVM parameter $\theta$, together with the theoretical predictions (red line).
%	\label{fig:NoiseFinal}}
%\end{figure}
%
%As seen from  Fig.\,\ref{fig:NoiseFinal}., the theoretical predictions of $N(\mathcal{M^\theta},\sigma_z)$ are reproduced evidently. In the next Section we move on to the entropic \emph{disturbance} $D(\mathcal{M}, B)$. However we will omit some steps, which are similar to the data treatment of the noise when evaluation the disturbance $D(\mathcal{M}, B)$ on the next pages.

\newpage

\subsection{II.2 Disturbance Measurement} \begin{figure}[!b]
	\includegraphics[scale=0.55]{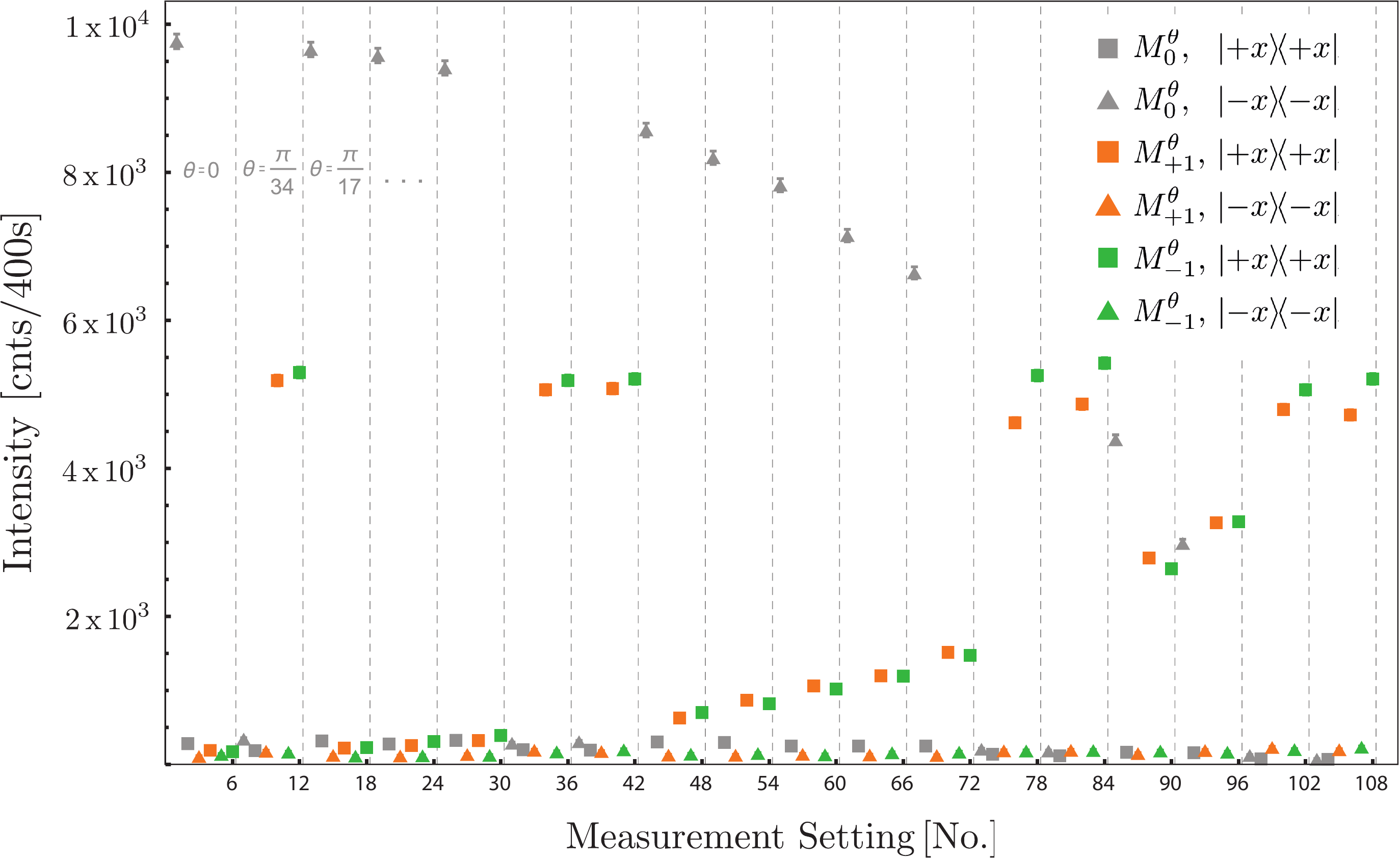}
	\caption{Raw data $I^b_{m.b'}$ (with $m=-1,0,+1$ and $b'=\pm 1$) of the \emph{disturbance} measurement $D_\mathcal E(M^\theta,\sigma_x)$ of the \textcolor{black}{three-outcome} POVM $M^\theta$ and projective $B=\sigma_x$ measurement for 400 seconds.
	\label{fig:DataDist}}
\end{figure}
\begin{figure}[!b]
	\includegraphics[scale=1.27]{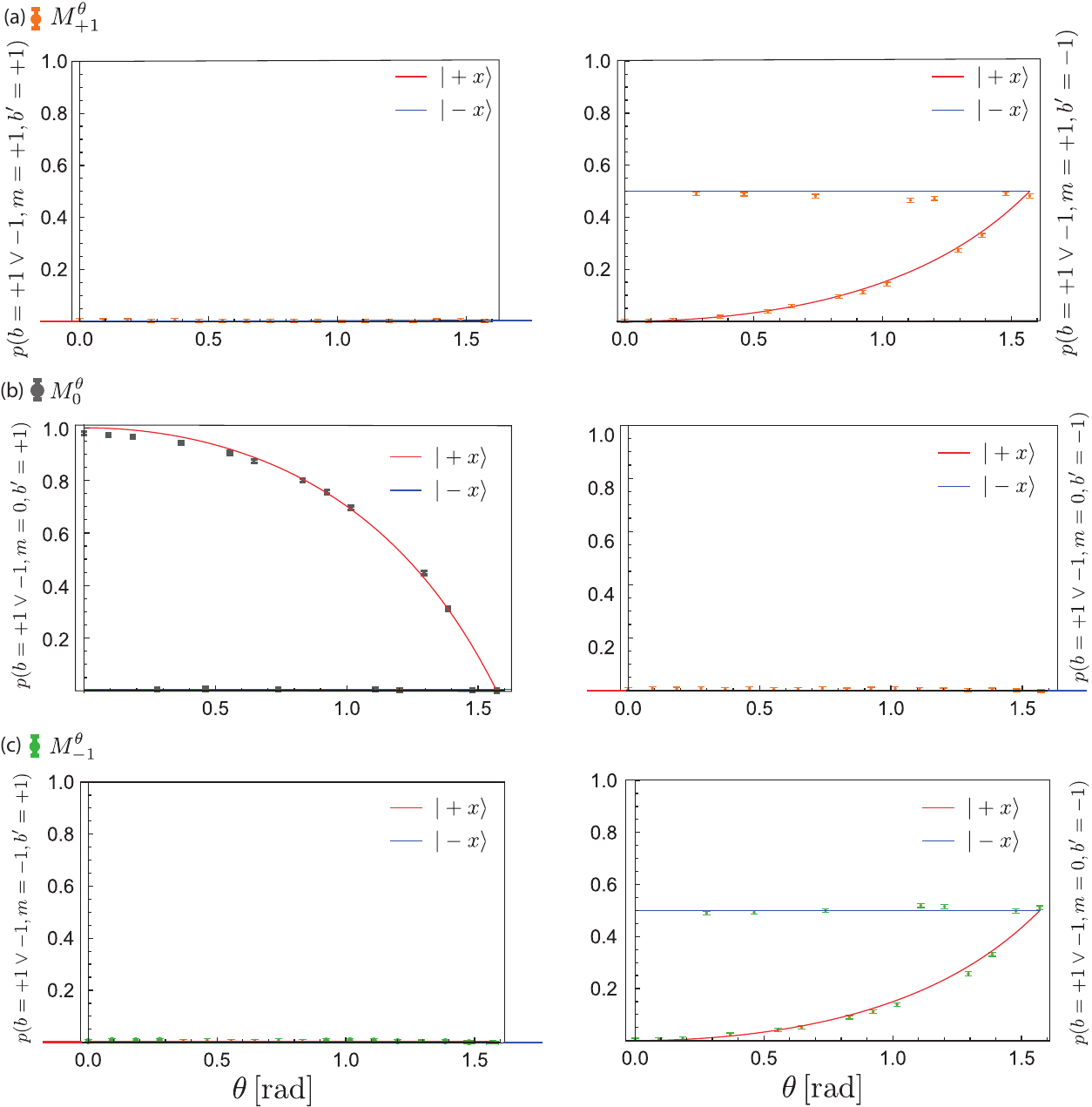}
	\caption{Normalized data of $M_{+1}^\theta$ (a), $M_{0}^\theta$ (b), and $M_{-1}^\theta$ (c), with "indefinite" input state ($\ket{ +x}$ or $\ket {-x}$) split up in the two output channels of the subsequent projective $B$ measurement with $B=\sigma_x$.
	\label{fig:pBBS}}
\end{figure}
For the disturbance measurement $D_\mathcal E(\mathcal M^\theta,B)$ the \textcolor{black}{three-outcome} POVM measurement is followed by a subsequent projective measurement of an observable $B=\sigma_x$. In addition, an \textcolor{black} {optimal correction} operation \textcolor{black}{$\mathcal E_m^{\rm{opt}}$ }in between the two measurements maps $\boldsymbol {n}_{-1}$ and $\boldsymbol {n}_{1}$ onto the negative $x$-axis and $\boldsymbol {n}_{0}$ onto the positive $x$-axis, respectively.
%
%\begin{figure*}[!h]
%	\includegraphics[width=0.8\textwidth]{Images/SchemeNewDistV4}
%	\caption{Scheme for disturbance, $D_\mathcal E(\mathcal M,B)$ measurement.
%	\label{fig:distscheme}}
%\end{figure*}
%
Uniformly distributed eigenstates of the observable $B$, denoted as $\{\ket{b_i}\}=\{\ket {+x},\ket{-x}\}$ \textcolor{black}{and }associated with random variable $\mathbb{B}$, are fed to the same instrument $\mathcal{M}^\theta$. Due to the disturbing nature of the measurement apparatus $\mathcal{M^\theta}$, generally, a loss of correlation occurs. The correlation between the eigenvalue $b$ corresponding to the state prepared and the outcome $b'$ of the second now \emph{projective} measured, which will be used to define the \emph{disturbance}, is characterized by the joint probability distribution $p(b,b')$, allowing to calculate the disturbance $D_{\mathcal E}(\mathcal M^\theta,B)$ using Eq.(\ref{eqS:DistDefn}). 
%\begin{figure}[!h]
%	\includegraphics[width=0.96\textwidth]{Images/SetupV3simple}
%	\caption{Neutron polarimetric setup for \emph{noise-disturbance} measurement $D_{\mathcal{E}}(\mathcal{M}, B)$ of the three output POVM $M^\theta$. DC coils 3 and 4 were redesigned to account for beam displacement $\Delta y$.
%	\label{fig:setup2}}
%\end{figure}
%

In the actual experiment, the detector (Boron trifluoride cylindric count tube, diameter \o=3 inch, active volume of length $L_{\rm{act}}=30$\,cm) was placed \emph{horizontally, transversal} to the beam. This was done to account for the beam displacement $\Delta y\sim 10\,$mm, caused by the \textcolor{black}{tilt} of the second supermirror, when setting the POVM weights. With this configuration a maximal count rate $I_{\rm{max}}=25$  cnts/sec is recorded. 
During the measurement the POVM parameter $\theta$ is varied between $\pi/2$ and 0 in steps of $\pi/34$. For each value of $\theta$ now \emph{six} intensities $I^b_{m,b'}$, belonging to the $+b$ and $-b$ measurement of the POVM outputs $M_0^\theta$,  $M_{+1}^\theta$, and $M_{-1}^\theta$, are recorded in a measurement time $t_{\rm{meas}}=400$ seconds, which is plotted in Fig.\,\ref{fig:DataDist} (for higher statistics also a second data set with  $t_{\rm{meas}}=800$ seconds was recorded). For each value of $\theta$ an initial state (eigenstate of $B=\sigma_x$) is chosen by \textcolor{black}{a} random generator. Again, the result is \emph{blinded} during the measurement but stored in file for a later comparison with the obtained values for the disturbance $D_{\mathcal E}(\mathcal M^\theta,B)$. The following sequence was randomly generated:  

\begin{eqnarray*}
& \{\theta=0,\vert +x\rangle\},  \{\theta=\frac{\pi}{34},\vert -x\rangle\}, \{\theta=\frac{\pi}{17},\vert +x\rangle\},  \{\theta=\frac{3\pi}{34},\vert +x\rangle\},  \{\theta=\frac{2\pi}{17},\vert +x\rangle\}, \{\theta=\frac{5\pi}{34},\vert -x\rangle\}, \{\theta=\frac{3\pi}{17} ,\vert -x\rangle\} \nonumber\\
&   \{\theta=\frac{7\pi}{34},\vert +x\rangle\},    \{\theta=\frac{4\pi}{17},\vert +x\rangle\},      \{\theta=\frac{9\pi}{34},\vert +x\rangle\},     \{\theta=\frac{5\pi}{17},\vert +x\rangle\},   \{\theta=\frac{11\pi}{34},\vert +x\rangle\},  
 \{\theta=\frac{6\pi}{17},\vert -x\rangle\},   \{\theta=\frac{13\pi}{34},\vert -x\rangle\}   \nonumber\\  &   
 \{\theta=\frac{7\pi}{17},\vert +x\rangle\},  \{\theta=\frac{15\pi}{34},\vert +x\rangle\},  \{ \theta=\frac{8\pi}{17},\vert -x\rangle\},   \{\theta=\frac{\pi}{2},\vert -x\rangle\}  
  \end{eqnarray*}
As before in the noise measurement, the count rates are detangled \textcolor{black}{according} to their corresponding $B$ measurement and POVM output.

 Next a background correction is applied, \textcolor{black}{by subtracting} the background counts of $I^{b'}_{\rm{bg}}=0.176\pm 0.008\,$ cnts per sec resulting in the intensity $I^{b'}_{\rm{bgCorr}}(M_{m}^\theta)$ and a overall contrast of $C=0.97$ is taken into account. Following the same procedure as for the noise, the count rates are normalized (equipped with statistical and systematical error) by the total number of counts which gives the $six$ probabilities $p(b=+1\vee -1,m,b')$ with $m=-1,0,1$ and $b'=\pm 1$, which is plotted in Fig.\,\ref{fig:pBBS} (a), (b) and (c), left and right, respectively.

\begin{figure}[!t]
	\includegraphics[scale=0.62]{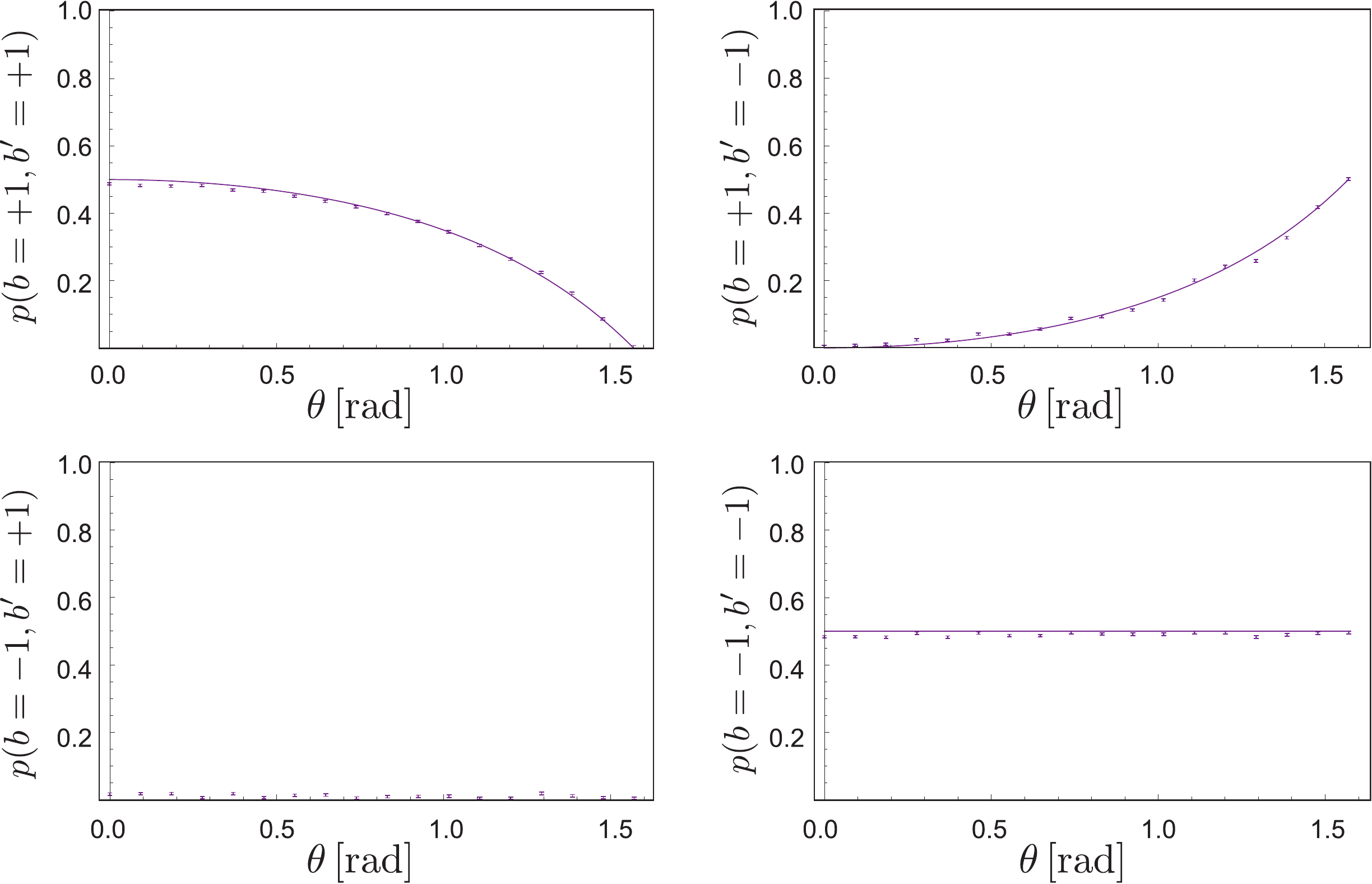}
	\caption{Joint probabilities $p(b,b')$ with $b=\pm 1$ and $b'=\pm 1$, together with theoretical predictions.
	\label{fig:JointBBS}}
\end{figure}

Again the data points are separated according to the input state 
$\ket {+x} \rightarrow b=+1$ and  $\ket {-x} \rightarrow b=-1$, which gives in total 12 probabilities $p(m,b,b')$ with  $m=-1,0,1$,  $b=\pm 1$ and $b'=\pm 1$ (not shown here). Since the disturbance is defined as 
\begin{eqnarray}
D_{\mathcal{E}}(\mathcal{M}^\theta, \sigma_x)= -\sum_{b,b'}p(b, b') \log_2p(b| b')=-\sum_{b,b'}p(b, b') \log_2\frac{p(b,b')}{p(b')},
   \end{eqnarray}
   we \textcolor{black}{have} to calculate the joint probability $p(b,b')$ via $p(b,b')=\sum_{m=-1}^1 p(m,b,b')$, which is plotted in Fig.\,\ref{fig:JointBBS}. The theoretical curves are given by 
\begin{eqnarray}      
   p(b,b')=\frac{1-b'+(1+b'+2bb')\cos\theta}{4(1+\cos\theta)},
    \end{eqnarray} 
\textcolor{black}{plotted as} purple lines in Fig.\,\ref{fig:JointBBS}.

\begin{figure}[!h]
	\includegraphics[scale=0.62]{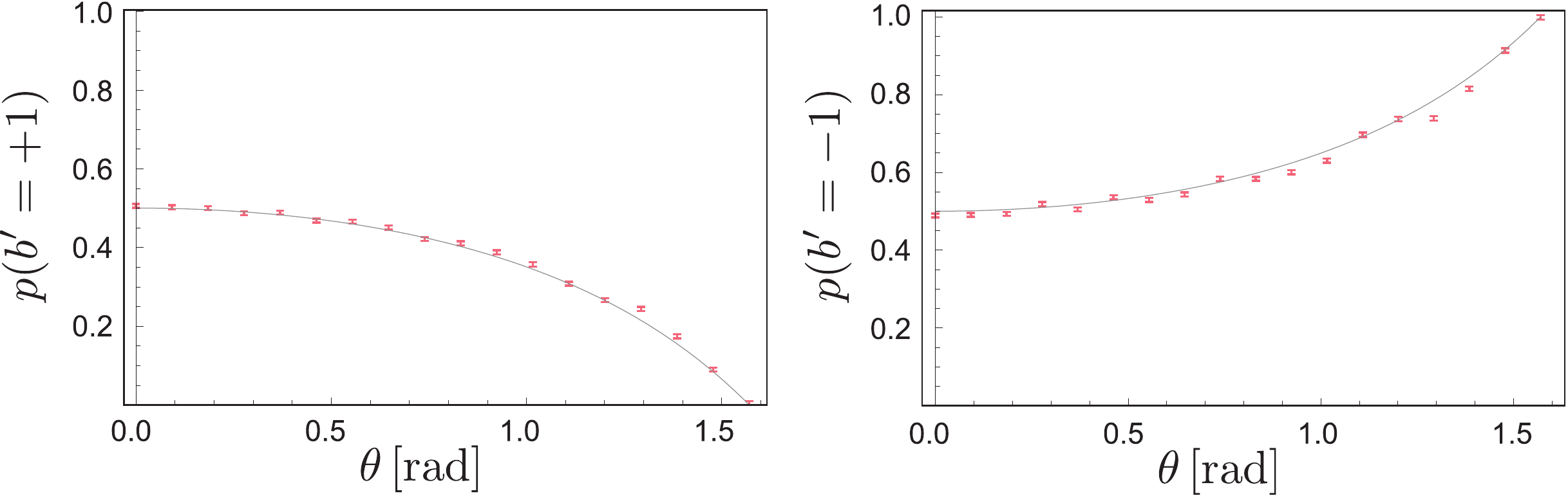}
	\caption{Marginal probabilities $p(b')$, together with theoretical predictions.
	\label{fig:MarginalsBs}}
\end{figure}

Next we \textcolor{black}{calculate the} marginal probabilities $p(b')$ by summation of the data from above. The theoretical curves are given by 
\begin{eqnarray}      
p(b')=\sum_b p(b,b')=\frac{1-b'+\cos\theta+b'\cos\theta}{2+2\cos\theta},
    \end{eqnarray} 
\textcolor{black}{plotted as} gray lines in Fig.\,\ref{fig:MarginalsBs}.

\begin{figure}[!t]
	\includegraphics[scale=0.4]{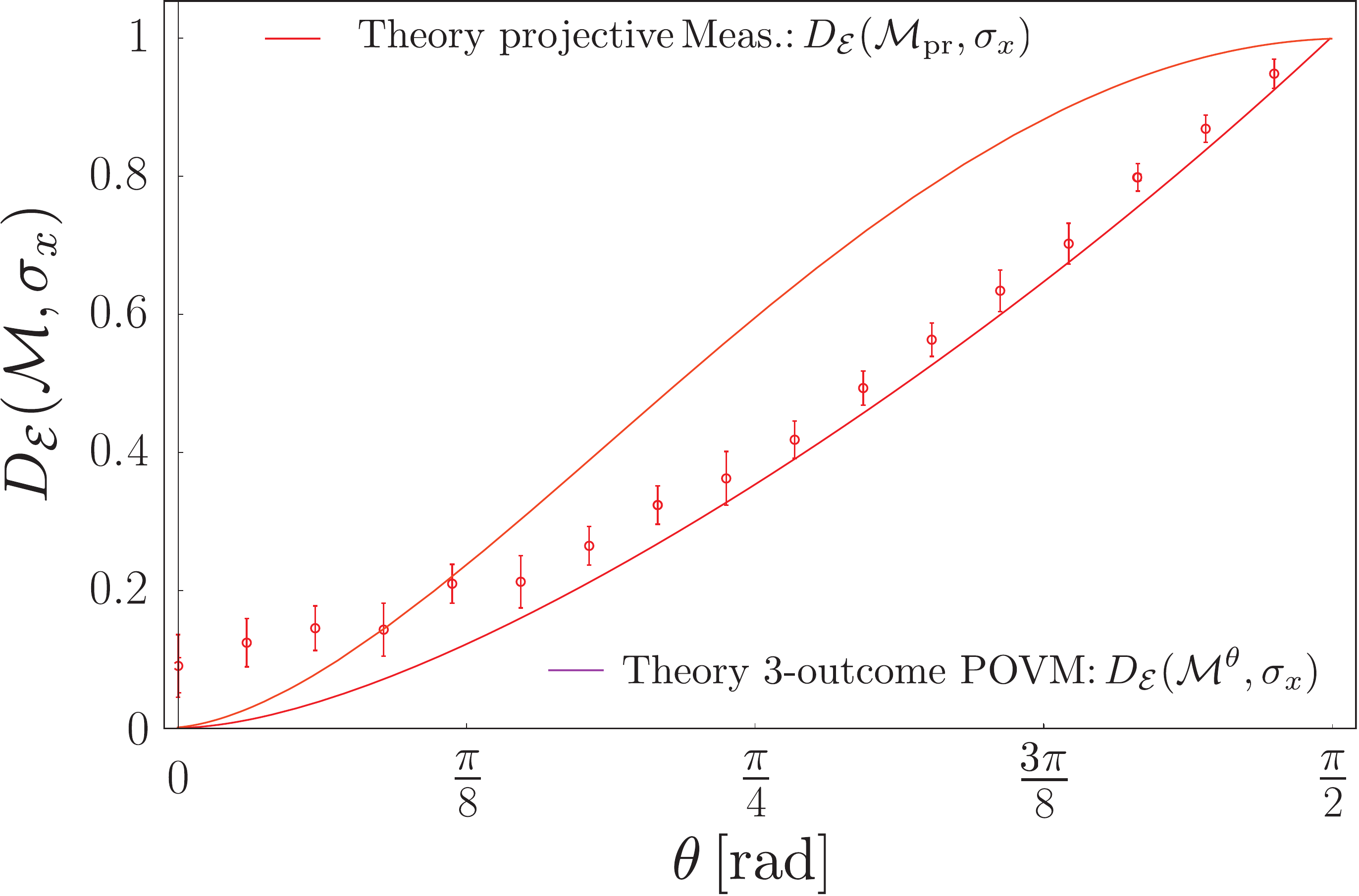}
	\caption{Plot of the noise $N(\mathcal M^\theta,\sigma_z)$ of the \textcolor{black}{three-outcome} POVM $M^\theta$ as a function of the POVM parameter $\theta$, together with the theoretical predictions for the three-outcome POVM (\textcolor{black}{red} line) and projective measurement (orange). \textcolor{black}{Error bars correspond to plus/minus one standard deviation.}
	\label{figSupp:DistFinal}}
\end{figure}

Finally, the disturbance is calculated applying the four joint probabilities $p(b,b')$ from above via the conditional entropy $H(\mathbb{B}|\mathbb{B}')$, $D_{\mathcal{E}}(\mathcal{M}^\theta, B) := H(\mathbb{B}|\mathbb{B}') = -\sum_{b,b'}p(b, b') \log_2 p(b| b')=-\sum_{b,b'}p(b, b') \log_2\frac{p(b,b')}{p(b')}$,
which is depicted in Fig.\,\ref{figSupp:DistFinal}. See also Fig.\,5 of the main text, where the disturbance $D_{\mathcal{E}}(\mathcal{M}^\theta, B)$ is plotted versus the noise $N(\mathcal M^\theta,A)$ with $A=\sigma_z$ and $B=\sigma_x$.

%A parametric plot of the experimental results of the noise-disturbance measurement is given in Fig.\,\ref{figSupp:NoiseDistFinal}, where the disturbance $D_{\mathcal{E}}({\mathcal M^\theta}, \sigma_x)$ is plotted versus the noise $N(\mathcal M^\theta,\sigma_z)$. 

%\begin{figure}[!h]
%\begin{center}
%	\includegraphics[scale=0.45]{ImagesSupp/ND9Supp}	\caption{\textcolor{blue}{NOTE: REWRITE CAPTION}Experimental comparison between noise-disturbance plot for \textcolor{red}{successive} projective measurements $N(\mathcal{M_{\rm{pr}}},\sigma_z)$ vs. $D_{\mathcal{E}}(\mathcal{M_{\rm{pr}}},\sigma_x)$  (green) - taken from  \cite{Sulyok15} - together with theoretical predictions in red and $N(\mathcal M^\theta,\sigma_z)$ vs. $D_\mathcal{E}(\mathcal M^\theta,\sigma_x)$ (blue)  for the \textcolor{red}{three-outcome} POVM $M^\theta$ of the measurement apparatus $\mathcal M^\theta$, \textcolor{red}{with theory} in purple. \textcolor{red}{Error bars correspond to plus/minus one standard deviation}.
%	\label{figSupp:NoiseDistFinal}}
%\end{center}
%\end{figure}

\end{document}